\documentclass[article,fleqn,usenatbib]{mnras}
 \pdfoutput=1
 \usepackage{newtxtext,newtxmath}

\usepackage[T1]{fontenc}
\usepackage{ae,aecompl}

\usepackage{graphicx}	
\usepackage{amsmath}	
\usepackage{amssymb}	

\usepackage{booktabs}
\usepackage{multicol}
\usepackage{placeins}
\usepackage{tikz}
\usepackage{float}
 
\title[Scattering of cometary dust analogs]{Experimental phase  function and degree of linear polarization of cometary dust analogs}

\author[E. Frattin et al.]{E. Frattin,$^{1,2}$\thanks{E-mail: elisa.frattin@gmail.com}
 O. Mu\~{n}oz,$^{3,4}$
 F. Moreno,$^{3}$
J. Nava,$^{5}$
J. Escobar-Cerezo,$^{6}$\newauthor
J. C. Gomez Martin,$^{3}$ 
 D. Guirado,$^{3}$
 A. Cellino,$^{7}$
 P. Coll,$^{8}$
 F. Raulin,$^{8}$
I. Bertini,$^{1}$\newauthor
 G. Cremonese,$^{2}$
 M. Lazzarin,$^{1}$  
G. Naletto, $^{9,10,11}$ 
F. La Forgia $^{1}$.
  \\      
    $^{1}$     University of Padova, Department of Physics and Astronomy
"G. Galilei", Vicolo dell'Osservatorio 3, 35122 Padova, Italy\\
 $^{2}$   INAF, Astronomical Observatory of Padova, Vicolo dell'Osservatorio 5, 35122 Padova, Italy\\
 $^{3}$  Instituto de Astrof\'{\i}sica de Andaluc\'{\i}a, CSIC, Glorieta de la Astronomia s/n, E-18008 Granada, Spain\\
 $^{4}$  Advanced Optical Imaging Group, School of Physics,
 University College Dublin, Dublin, Ireland  \\  
  $^{5}$   University of Padova, Department of Geosciences, Via G.Gradenigo 6, 35131 Padova, Italy\\
  $^{6}$ Department of Physics, University of Helsinki, Finland\\
$^{7}$  INAF-Astrophysical Observatory of Torino, strada Osservatorio 20, Pino Torinese (TO)\\
 $^{8}$   Laboratoire Inter-Universitaire des Systemes Atmospheriques, Universites Paris 12-Paris 7, CNRS\\
  $^{9}$  University of Padova, Department of Physics and Astronomy "G. Galilei", Via Marzolo 8, 35131 Padova, Italy\\
  $^{10}$ University of Padova, Center of Studies and Activities 
for Space (CISAS) "G. Colombo", Via Venezia 15, 35131 Padova, Italy\\
  $^{11}$ CNR-IFN UOS Padova LUXOR, Via Trasea 7, 35131 Padova, Italy
  }
\date{Accepted 2018 December 22. Received 2018 December 20; in original form 2018 July 19.}

\pubyear{2018}
\begin{document}
\maketitle
\begin{abstract}
We present experimental phase function and degree of linear polarization curves for seven samples of cometary dust analogues namely: ground pieces 
of Allende, DaG521, FRO95002 and FRO99040 meteorites, Mg-rich olivine and pyroxene, and a sample of organic tholins. The experimental curves have 
been obtained at the IAA Cosmic Dust Laboratory at a wavelength of 520 nm covering a phase angle range from 3$^{\circ}$ to 175$^{\circ}$. We also provide values of the backscattering enhancement (BCE) for our cometary 
analogue samples.  The final goal of this work is to compare our experimental curves with observational data of comets and asteroids to better constrain the nature of cometary and asteroidal dust grains. 
All measured phase functions present the typical behavior for $\mu$m-sized
cosmic dust grains.  Direct comparison with data provided by the OSIRIS/Rosetta camera for comet 67P Churyumov-Gerasimenko reveals significant differences and supports the idea of a coma dominated by big 
chunks, larger than one micrometer. The polarization curves are qualitatively similar to ground-based observations of comets and 
asteroids. The position of the inversion polarization angle seems to be 
dependent on the composition of the grains. We find opposite dependence of the maximum of the polarization curve for grains sizes in the Rayleigh-resonance and geometric optics domains, respectively. 
\end{abstract}


\section{Introduction}

Dust is a fundamental constituent of planetary systems in all their stages of development. For example, the origin of planetary systems involves a phase of aggregation of submicron-sized dust grains  by streaming instability into protoplanetary disks \citep{johansen,blum,blum_review}. In our mature Solar System, dust can be found in the interplanetary medium orbiting around the Sun as a product of the disintegration of comets and asteroidal collisional cascading (the Zodiacal Cloud), as well as in the rings of the giant planets. In addition, dust is present in the regolith surfaces of planets, asteroids and comets, and in dense atmospheres (e.g. Venus, the Earth, Mars, Titan) in the form of aerosol. Because of the ubiquity of dust and its involvement in countless physical phenomena, from atmospheric radiative transfer to cometary activity through the tracing of Solar System   primitive materials, it is important to study its nature and properties.\\
Comets are the most appropriate targets to study the early Solar System, since they are among the most pristine objects orbiting the Sun. 
Because of their low-velocity, gentle accretion process, some of their most primitive constituents (the grains) have remained intact, i.e. as they were when the comets formed \citep{fulle_pebbles}. During their passage at perihelion, these objects release refractory materials driven by the sublimation of trapped  ices, generating a bright dust coma dominated by $\mu m$-sized particles composed of silicate minerals and organics. 
Comets unveil the composition of the primordial material from which they accreted, namely material located in the outer regions of the Solar System,  and put some constraints to the processes involved in its origin.
The flyby missions that first revealed the nature of cometary dust with in-situ measurements were the \textit{Giotto} \citep{reinhard_giotto} and \textit{Vega} \citep{sagdeev_vega} missions to 1P/Halley in 1986. The 'Halley Armada' was followed up by the Deep Space 1 mission to comet 19P/Borrelly in 2001 \citep{soderblom_borrelly}. A major breakthrough came with the \textit{Stardust} mission flyby of comet 81P/Wild in 2004 \citep{horz_stardust}, and subsequent return to Earth of samples of dust collected directly from the coma of the comet. 
Later, missions to comet 9P/Temple in 2005 (Deep Impact) 2005 \citep{deep_impact_mission} and to comet 103P/Hartley in 2010 (EPOXI) \citep{epoxi_mission} provided new insights of these two objects.
More recently, the Rosetta mission has been able, for the first time, to follow the trajectory of the cometary nucleus and observe its evolution during its trajectory before and after perihelion. The coma of comet 67P/Churyumov-Gerasimenko (hereafter 67P) was observed during two entire years (2014-2016) from the very inside, reaching distances of few km from the nucleus. The astonishing images obtained by the OSIRIS instrument  on board the spacecraft have shed new light into the properties of cometary dust \citep{sierks2015nature} . There is now evidence  that  67P formed by gentle accretion of pebbles smaller than 1 cm  at velocities of the order of 1 m/s \citep{fulle-blum}. Moreover, photometric analysis of the overall coma and single grains reveal that the majority of the material has similar spectral properties to those of nucleus surface \citep{frattin}, which enables using the dust coma as a proxy indicator of nucleus properties.\\
The sunlight incident on a cometary dust envelope is partly absorbed and partly scattered by the particle cloud.
Spacecraft and ground based observations of scattered light show characteristics strictly dependent on the nature and physical properties (composition, size distribution, shape, roughness, etc) of the material composing the coma. Since each material has characteristic scattering signatures, scattering theory can be used to interpret observations and retrieve from them some of the properties of the dust. Experimental data of the angular distribution of the scattered intensity and degree of linear polarization of clouds of cosmic dust analogs assist in the interpretation of in-situ and remote sensing observations. Since these quantities are intimately related to the nature of the particles, laboratory data are used as a reference for comparison and interpretation, allowing a correct analysis of the observational data \citep{munoz_allende, volten_forsterite, munoz_mm-size}. Modeling of the interaction between electromagnetic radiation and particles in turn, supports the experiments, helping to discriminate unambiguously dust features corresponding to specific observables \citep{liu_gaussian_sphere, zubko_enstatite, jesus_largeIDP}. All these different approaches (observations, laboratory experiments and theoretical  simulations) provide essential contributions to the investigation of the behavior of dust particles as radiation scatterers.\\

In this work we present experimental phase function and degree of linear polarization curves as functions of the observational phase angle of a selected set of cometary dust analogs: four meteoritic samples, two minerals (Mg-rich olivine and pyroxene), and a sample of organic particles (tholins). Light scattering measurement have been performed at the Cosmic Dust Laboratory (CODULAB) \citep{munoz_lab} at the Instituto de Astrof\'{\i}sica de Andaluc\'{\i}a (IAA), spanning a phase angle range from 3$^{\circ}$ to 175$^{\circ}$ at a wavelength of 520 nm. We combine the new measurements presented in this work with the scattering matrices at 442 nm and 633 nm of a sample of the Allende meteorite and of size-segregated Mg-rich olivine samples that have been previously presented by \cite{munoz_allende}. Finally, we compare the experimental data with observations of asteroids and cometary dust envelops, with special emphasis on the observations of 67P by Rosetta, in order to put  constraints on the nature of cometary dust. 

  



\section{Light scattering theory}
A beam of quasi-monochromatic light is physically defined by the Stokes vector $\textbf{I}=\{I,Q,U,V\}$, where $I$ is proportional to the total flux of the light beam and $Q$ is related to the linear polarization of the light beam, representing the difference between the two components of the flux along the $x$-axis and the $y$-axis. $U$ is defined by the difference of the two components of the flux along the directions rotated by 45$^\circ$ from the $x$ and $y$ axis. $V$ is related to the circular polarization, defined as the difference between the left-handed and the right-handed polarized components of the flux.\\
The interaction between a light beam and a cloud of particles results in scattering of the incident light. The Stokes vector of the scattered beam $\textbf{I}_{sc}$
is related to the one of the incident beam $\textbf{I}_{in}$
by the so-called scattering matrix $\mathbf{F}$, through the following expression:
\begin{equation}
\begin{pmatrix}
I_{sc} \\ Q_{sc} \\ U_{sc} \\ V_{sc}\\
\end{pmatrix}
= \frac{\lambda^2}{4 \pi^2D^2} 
\begin{pmatrix}
F_{11} & F_{12} & F_{13} & F_{14} \\
F_{21} & F_{22} & F_{23} & F_{24} \\
F_{31} & F_{32} & F_{33} & F_{34} \\
F_{41} & F_{42} & F_{43} & F_{44} \\ 
\end{pmatrix}
\begin{pmatrix}
I_{in} \\ Q_{in} \\ U_{in} \\ V_{in}\\
\end{pmatrix},
\label{eq:Stokes}
\end{equation}
where $\lambda$ is the wavelength of the incident beam and $D$ is the distance of the detector from the particles.
The 16 dimensionless elements of the matrix depend on  physical characteristics of the particles such as their number, shape, dimension (through the size parameter $x=2\pi r/ \lambda$) and refractive index, as well as on  geometrical parameters such as their orientation in the space and scattering direction.\\ 
When the particles are randomly oriented all scattering planes are equivalent. Therefore, the scattering direction is fully described by the scattering angle, $\theta$, defined  by the angle between the directions of propagation of the incident and scattered beams. To facilitate direct comparison  with astronomical observations of comets and asteroids we use the phase angle, $\alpha =180^{\circ} -\theta$, throughout the text.  Furthermore, when the  particles present mirror symmetries, and are randomly oriented, the scattering matrix has 6 parameters \citep{hulst}:
\begin{equation}
\mathbf{F} = 
\begin{pmatrix}
F_{11} & F_{12} & 0 & 0 \\
F_{12} & F_{22} & 0& 0 \\
0 & 0 & F_{33} & F_{34} \\
0 & 0 & -F_{34} & F_{44} \\ 
\end{pmatrix}
\label{eq:simp_matrix}
\end{equation}

\begin{figure} 
    \centering
    \includegraphics[width =0.48\textwidth]{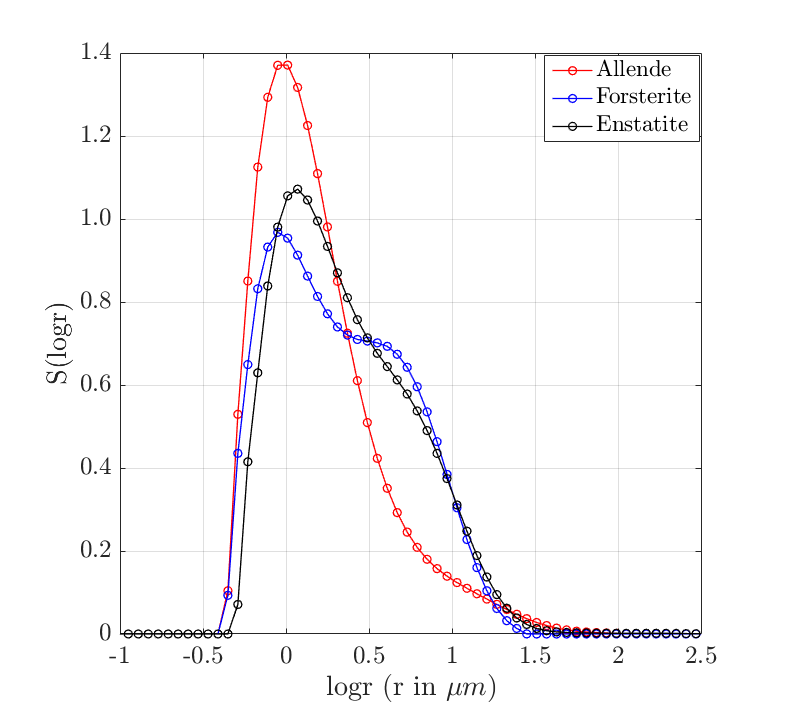}
    \caption{Size distribution of  samples of Allende meteorite, Forsterite and Enstatite.}
    \label{fig:sd_minerals}
\end{figure}

\section{Experimental apparatus}
The CODULAB facility is dedicated to measuring the elements of the scattering matrix (eq. \ref{eq:Stokes} and \ref{eq:simp_matrix}) as a function of the scattering angle of clouds of randomly oriented cosmic dust analogs. This is achieved by means of a gonio-nephelometer that has been described in detail previously by \cite{munoz_lab_2010}. In the experiments carried out for this work we employed an Argon-Krypton laser tuned to 520 nm as our light source. The laser beam passes through a polarizer and an electro-optic modulator before encountering a jet stream of randomly oriented particles produced by an aerosol generator \citep{munoz_lab}. The powder under analysis is contained in a cylindrical reservoir, and is pushed by a piston in a controlled manner towards a rotating brush, which disperses the particles into a flow of air which carries them trough a nozzle to the scattering volume. The particle density in the scattering volume, which is controlled by the speed of the piston, must be high enough to produce a detectable scattering signal, but low enough to avoid multiple scattering. In this way it better approximates the low density cometary coma.
The scattered light passes through optional optics before being detected by a photomultiplier, the \textit{detector}, which is mounted on a 1 m ring rail centered at the aerosol jet. In the measurements presented in this paper we have covered the phase angle range from 3$^{\circ}$ to 175$^{\circ}$ in steps of 5$^{\circ}$ within the  $10^{\circ}-20^{\circ}$ and $40^{\circ}-175^{\circ}$  ranges, and in steps of 1$^{\circ}$ within the $3^{\circ}-10^{\circ}$ and  $25^{\circ}-40^{\circ}$ ranges. Sinusoidal electro-optic modulation combined with lock-in detection allows determining all elements of the scattering matrix from eight different configurations of the optical components (eq. \ref{eq:Stokes})  and the assumption of reciprocity of the sample (see further details in \cite{munoz_lab_2010}). In order to keep the scattered intensity within the linearity range of the detector, a wheel of neutral density filters is placed between the laser head and the polarizer, which allows adapting the beam intensity to the scattering behavior of the aerosol sample at each particular angle. A second photomultiplier placed at a fixed angle, the \textit{monitor}, corrects for fluctuations in the aerosol beam. For each  position of the detector, 1000 measurements are conducted in about 2 seconds. The final value at each phase angle is obtained from the average of such measurements, with an associated error given by the standard deviation. Three measurement runs are carried out for each optical configuration in order to achieve further noise reduction. Due to the limited amount of material, we did not measure the complete scattering matrix but only the elements $F_{11}$ and $F_{12}$. For unpolarized incident light the $F_{11}$ is proportional to the scattered  flux. The measured values of the $F_{11}$ are  arbitrarily normalized so that they are equal to 1 at $\alpha=150^\circ$.  The $F_{11}$ normalized in this way is called the \textit{phase function}. The $-F_{12}/F_{11}$ ratio  (where the value of $F_{11}$ is not normalized) is equal to the degree of linear polarization, $P$, for unpolarized incident light,  expressed in terms of Stokes parameters as $P=-Q_{sc}/I_{sc}$.

\begin{figure} 
    \centering
    \includegraphics[width =0.48\textwidth]{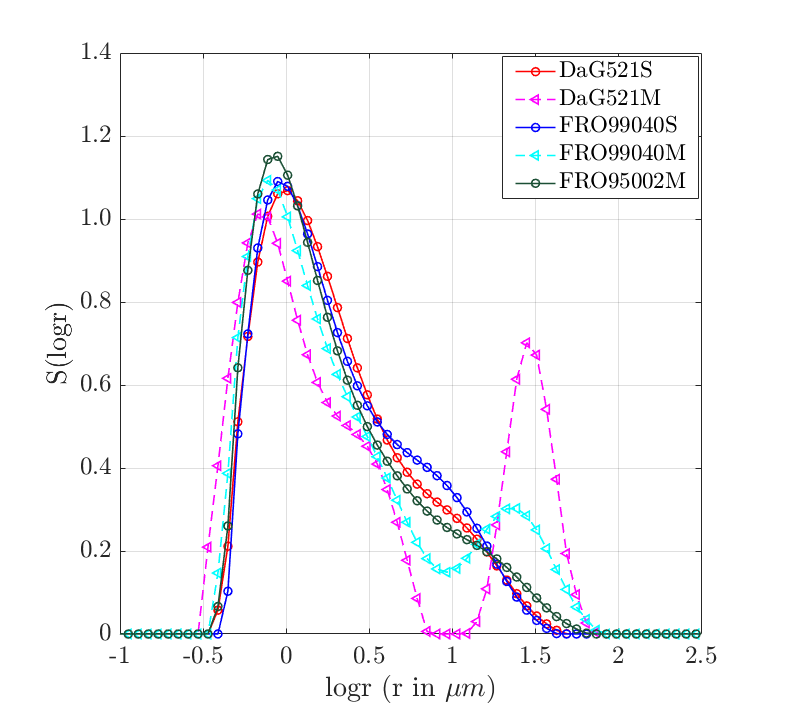}
    \caption{Size distribution of two different samples of DaG521, FRO95002 and FRO99040.}
    \label{fig:sd_meteorites_2}
\end{figure}

\begin{figure*}
\centering
\begin{tabular}{cc}
\includegraphics[width = 0.45\textwidth]{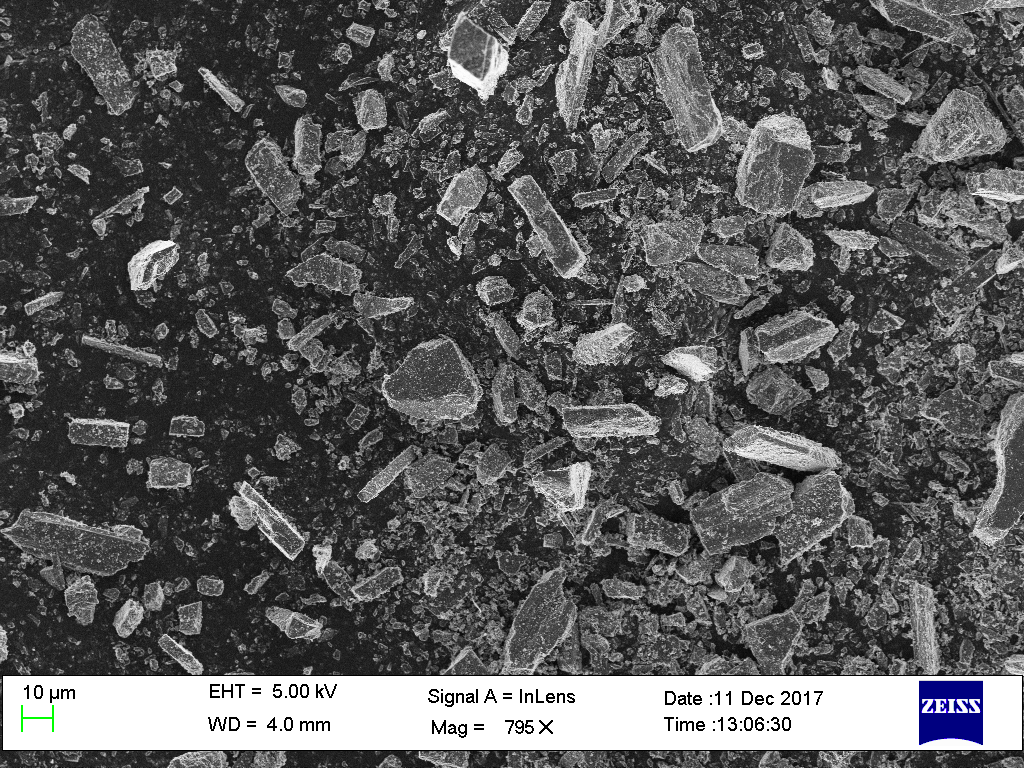}&
\includegraphics[width = 0.45\textwidth]{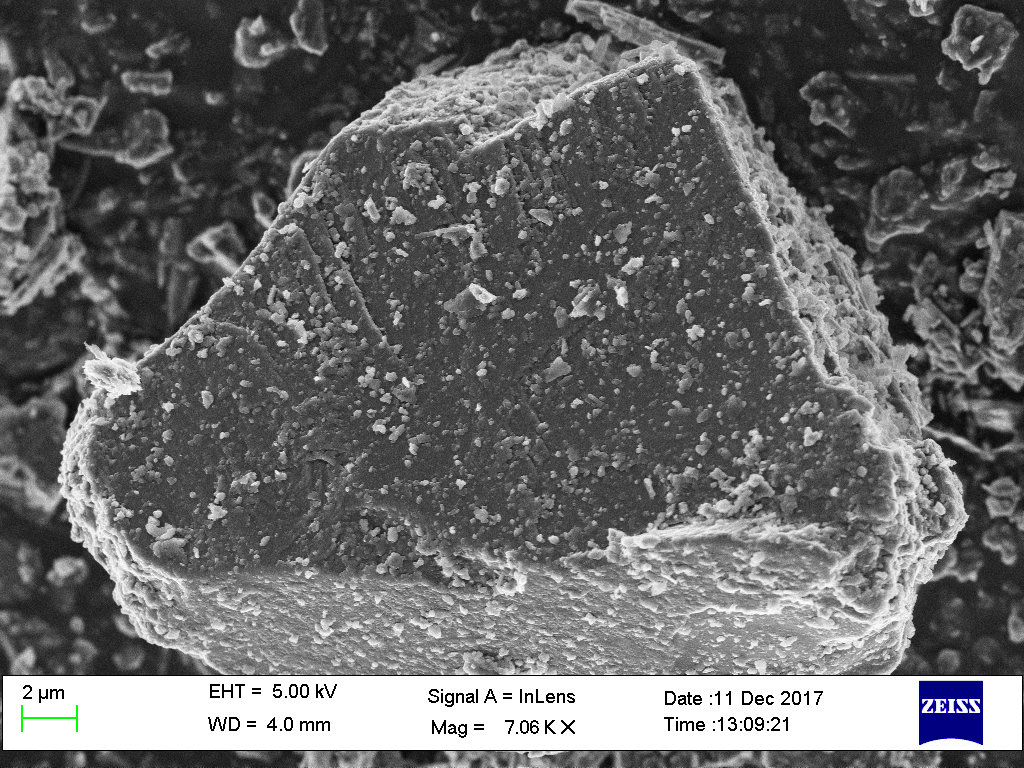}\\
(a) & (b)\\
\vspace{0.2cm}\\
\includegraphics[width = 0.45\textwidth]{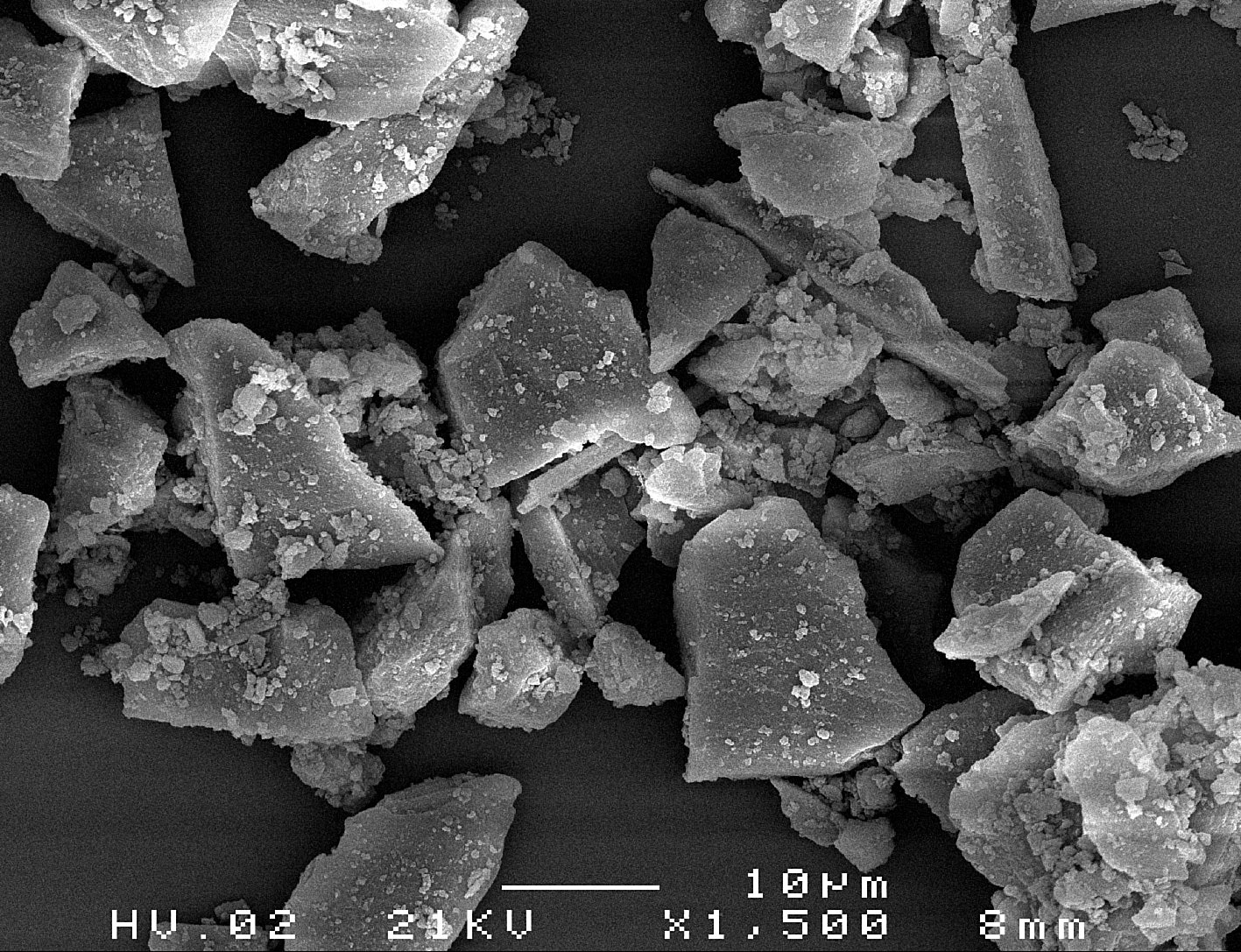}&
\includegraphics[width = 0.45\textwidth]{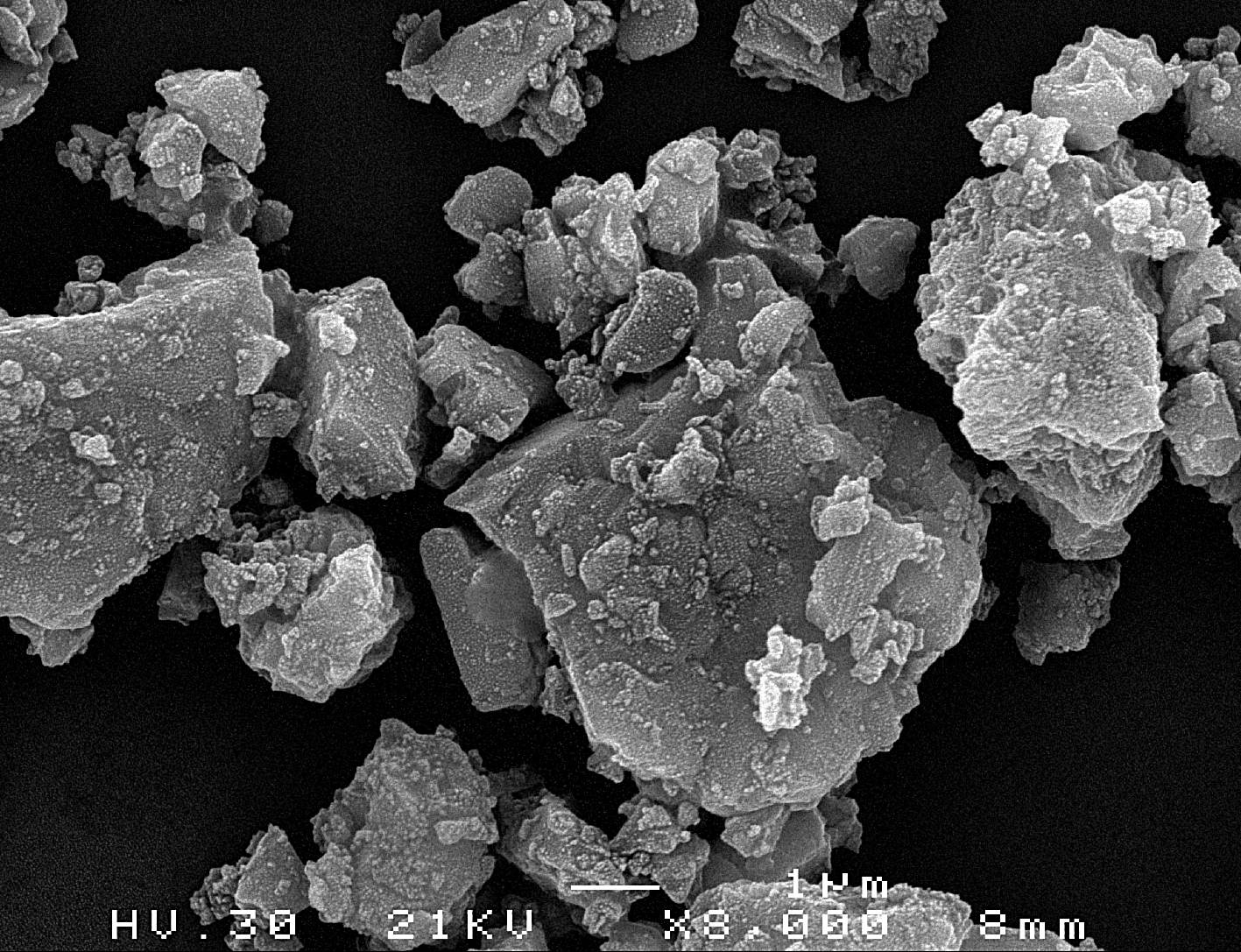}\\
(c) & (d)\\
\vspace{0.2cm}\\
\includegraphics[width = 0.45\textwidth]{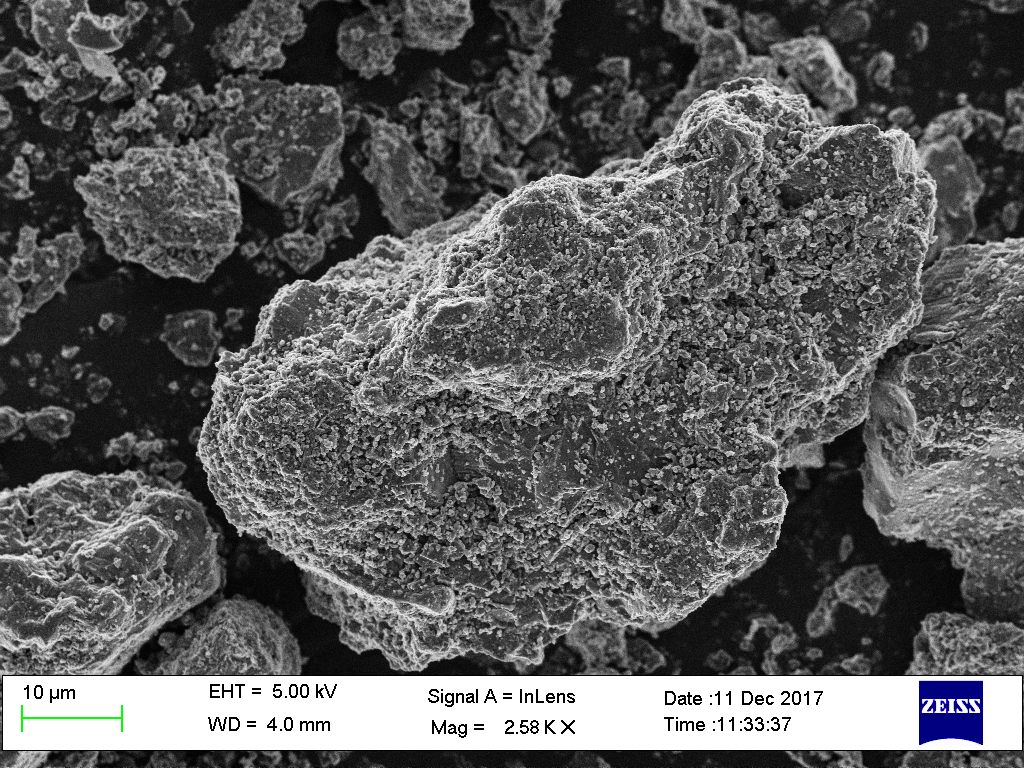}&
\includegraphics[width = 0.45\textwidth]{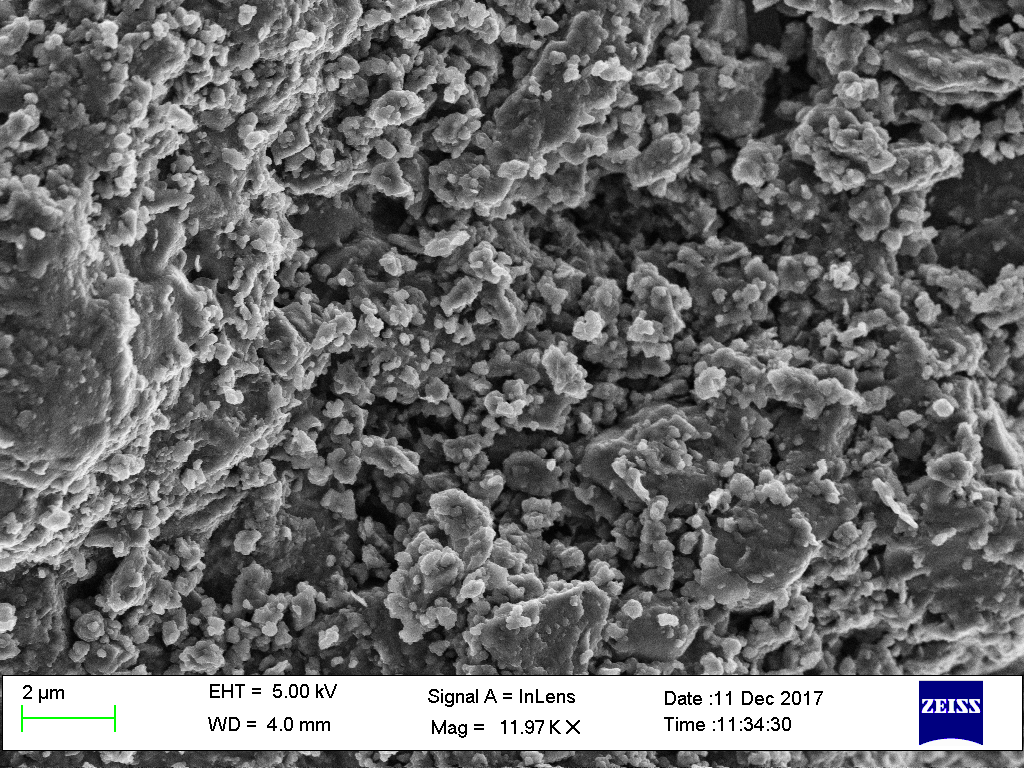}\\
(e) & (f)\\
\vspace{0.2cm}\\
\end{tabular}
 \end{figure*}
 
 \begin{figure*}
 \centering
\begin{tabular}{cc}
\includegraphics[width = 0.45\textwidth]{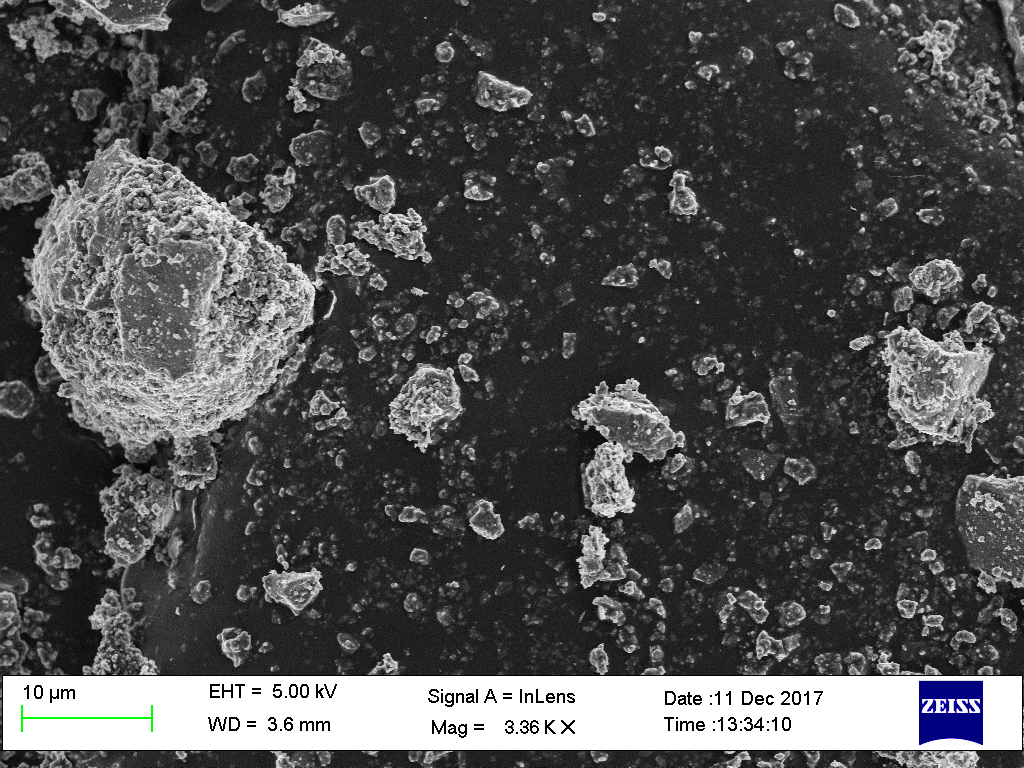} &
\includegraphics[width = 0.45\textwidth]{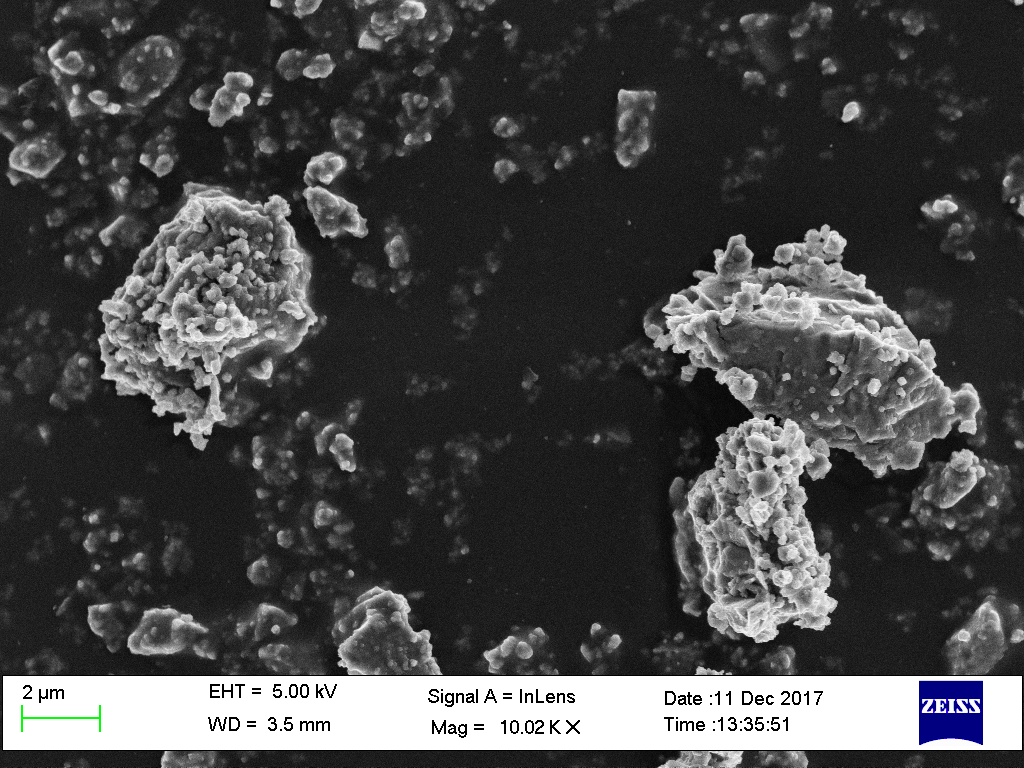}\\
(g) & (h)\\
\vspace{0.2cm}\\
\includegraphics[width = 0.45\textwidth]{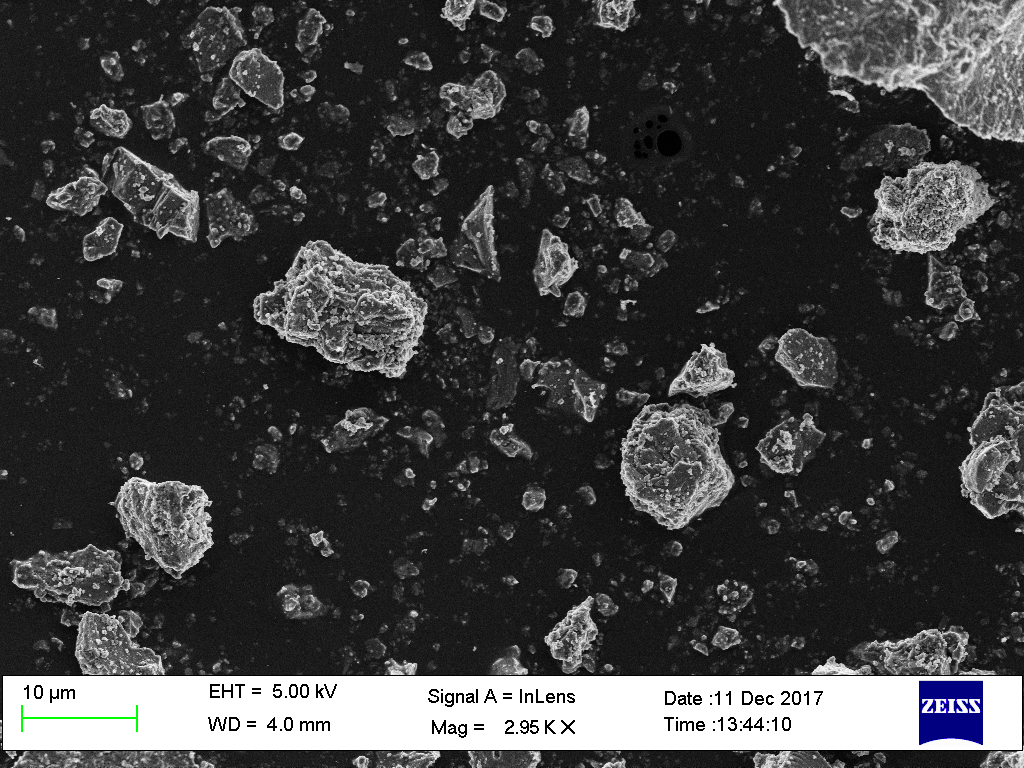}&
\includegraphics[width = 0.45\textwidth]{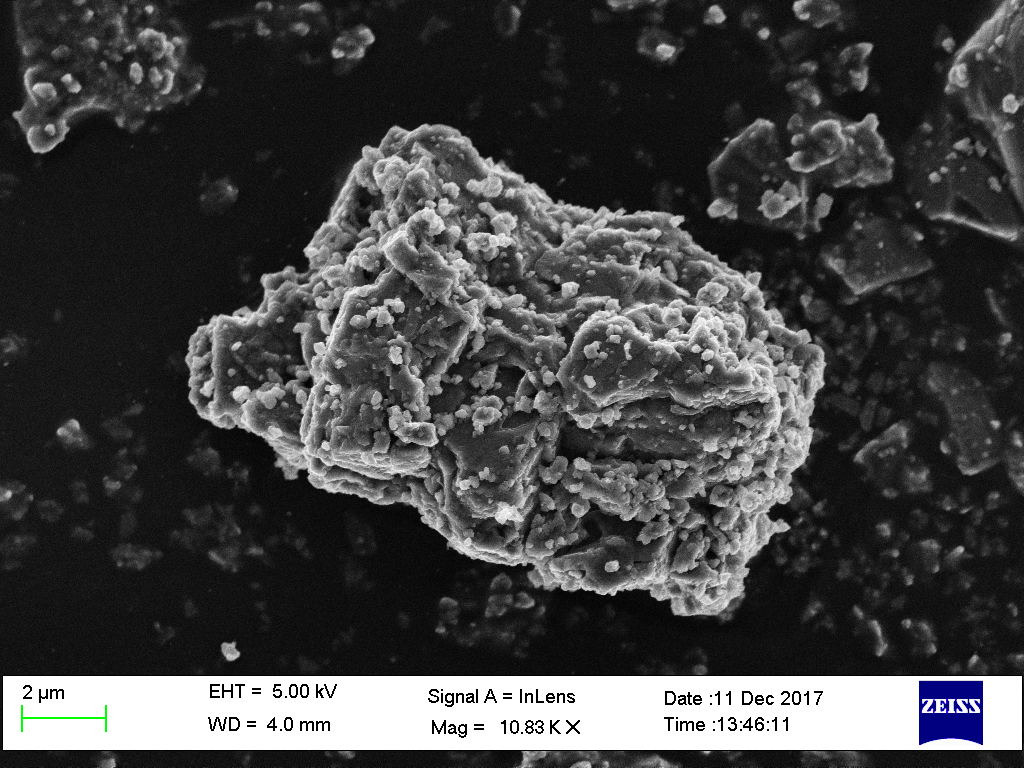}\\
(i) & (j)\\
\vspace{0.2cm}\\
\includegraphics[width = 0.45\textwidth]{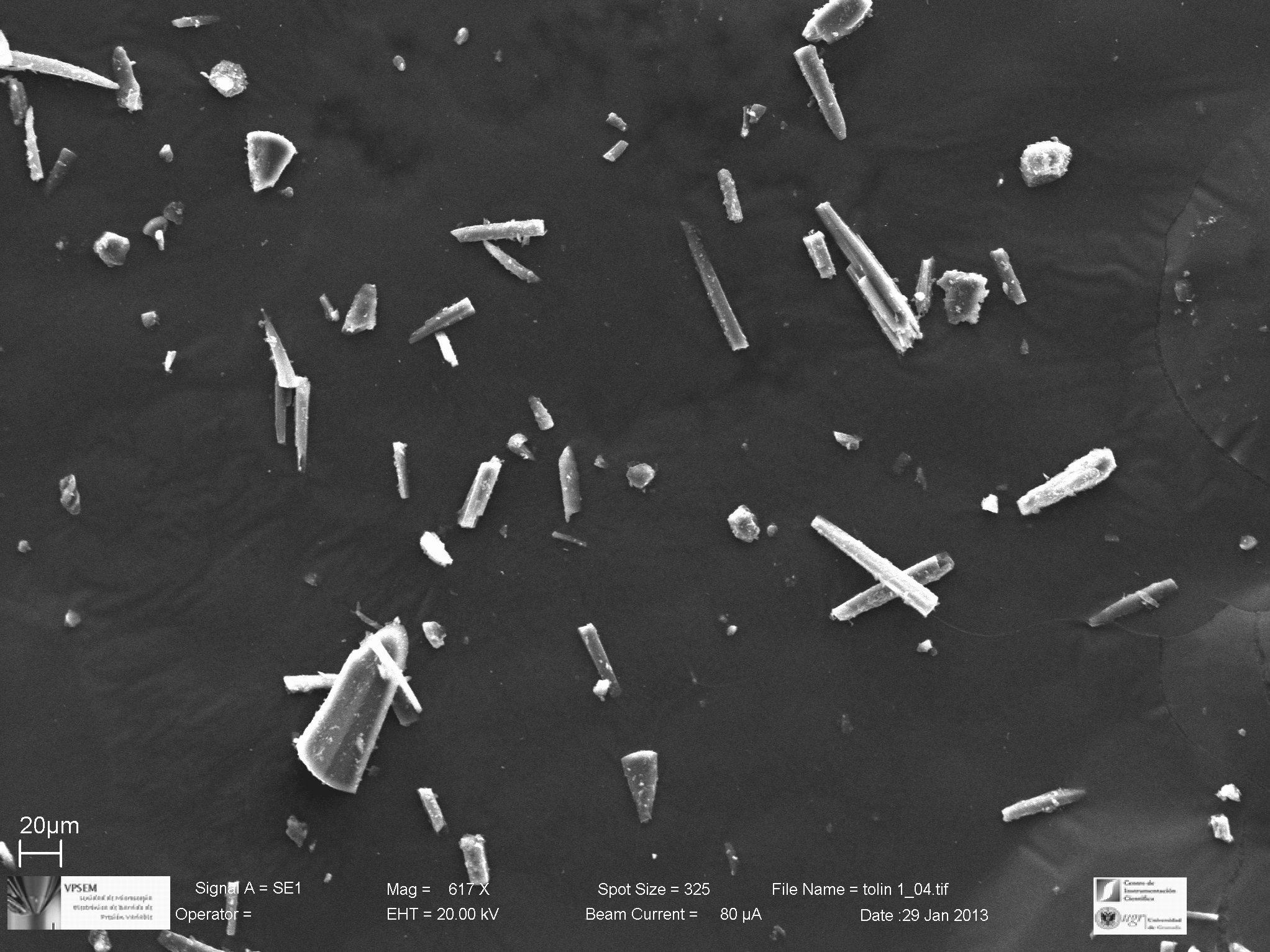}&
\includegraphics[width = 0.45\textwidth]{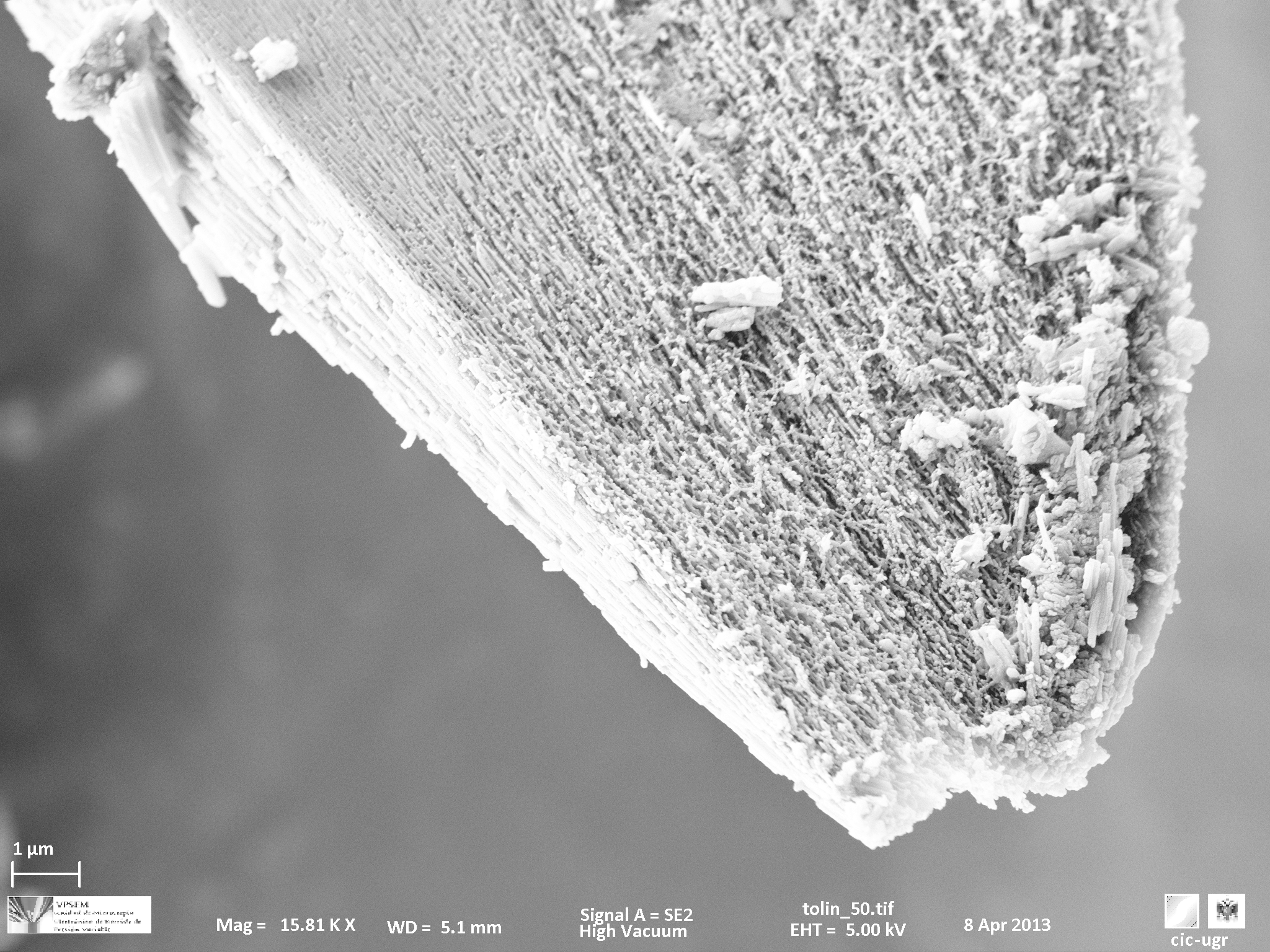}\\
(k) & (l)\\
\vspace{0.2cm}\\
\end{tabular}
  \caption{FESEM micro-photographs of enstatite (plates \textbf{a} and \textbf{b}),
forsterite (plate \textbf{c}), Allende (plate \textbf{d}),
DaG521 (plates \textbf{e} and \textbf{f}), FRO95002 (plates \textbf{g} and \textbf{h}),
FRO99040 (plates \textbf{i} and \textbf{j}) and tholins (plates \textbf{k} and
\textbf{l}).}
 \label{fig:sem_images}
\end{figure*}

\section{Materials}
We have selected four Carbonaceous Chondrites (CCs), two silicate minerals and one organic powder as sources of cometary dust analogs. Bulk samples of the meteorites DaG521, FRO95002 and FRO99040 were provided by the Museo dell'Antartide Felice  Ippolito, Siena especially for this work. Batches of particulate tholins were synthesized at the Laboratoire Inter-Universitaire des Systemes Atmospheriques (LISA), Paris. The Allende, olivine, and pyroxene samples belong to the IAA-CODULAB collection and have been used in a previous work \citep{munoz_allende}. Below we describe the bulk properties of these materials, the methods employed to obtain particulate samples suitable for our light scattering experiments, and the physical properties of the particles (optical constants, size distribution and morphology). 

\subsection{Bulk materials description}
\subsubsection{Meteorites}
Carbonaceous Chondrites meteorites  are among the oldest and most primitive materials in the Solar System. Age determination techniques indicate that they formed around 4.5 Gyr ago \citep{chen, manhes}.  
Since they are very old, these rocks suffered different primary processes, which record pre-accretionary histories in the solar nebula, and secondary processes like aqueous alteration, thermal metamorphism and shock metamorphism \citep{brearley}. \\
CCs are divided in several classes  based on composition: CI, CH, CO, CV, CK, CR.
Most of them are  fragments of primitive asteroids. They come from the NEO population, which  includes about 10\% of extinct comets, in addition to former asteroids \citep{demeo_binzel2008}. \\
The bulk composition of the Carbonaceous Ivuna (CI) group of chondrites is very close to the composition of the solar photosphere. Furthermore, interstellar grains that predate the Solar System formation are found in the matrix of these primitive chondrites \citep{brearley}.
CCs are characterized by chondrule sizes ranging from 1 mm in the CV group down to around 0.15 mm and 0.02 mm in the CO and CH groups, respectively. The CV and CO chondrites have the highest abundances of refractory inclusions, while CR and, above all, CH  chondrites have the highest content of metals  \citep{brearley, weisberg}. In Tab. \ref{tab:elements} we report the mean composition of CO and CV chondrites and of the  Allende meteorite \citep{metdb_database}.\\

\begin{table}
	\centering
	\begin{tabular}{cccc}
 &    CO & CV & Allende  \\
\hline\hline
Si & 15.8	& 16.0	& 16.0\\
 \hline
 Ti & 0.08 &	0.15 & 0.10 \\
 \hline
Al & 1.58 & 2.19 & 1.79\\
 \hline
Cr &  0.36 &	0.35 & 0.37\\
 \hline
Fe & 24.3 & 23.1 & 23.9 \\
 \hline
 Mn & 0.18 & 0.15 & 0.15\\
 \hline
Mg & 14.4 & 14.4 & 15.0   \\
 \hline
Ca & 1.62 & 2.5 & 1.89 \\
 \hline
K & 0.05 & 0.03 & 0.03 \\
 \hline
S & 1.94 & 1.63  & 1.77 \\
 \hline
P &  0.11 & 0.13 & 0.10\\
 \hline
Na &  0.35 & 0.35 & 0.33 \\
 \hline
Ni & 1.31 & 1.11 & 1.23\\
 \hline
 O & 37.9 & 37.9  &  37.2\\
\hline\hline\\
\end{tabular}
 	\caption{Mean composition of   CV and CO chondrites and of the Allende
 	meteorite in \%wt, taken from the meteorite database METDB \citep{metdb_database}.}
\label{tab:elements}
\end{table}


%
%

\paragraph*{DaG521 and Allende}

The meteorites DaG521 (Dar al Gani 521, found in Lybia in 1997 \citep{grossman1999}) and Allende (fallen near Parral, Chihuahua, Mexico, in 1969) 
are classified as CV3. CVs chondrites are characterized by large mm-sized chondrules, large refractory inclusions and abundant matrix (40 \%vol). They are divided into oxidized CV\textsubscript{OX} and reduced CV\textsubscript{RED} subgroups based on abundances of metal, magnetite and Ni content \citep{mcsween1}. The main component of CV matrices is Fe-rich olivine (fayalite Fa$\sim$30-60 and in some cases Fa$>$90), but there are substantial differences among the subgroups due to different late stage metasomatism and aqueous alteration \citep{krot1, krot2}. 
Other mineral constituents are Ca-Fe-pyroxene, andradite, Fe-Ni sulfides and magnetite.
Some chemical and textural features of the CV chondrites evidence a phase in which the parent body experienced temperatures <300$^{\circ}$C \citep{krot2}.\\
DaG521 is the only meteorite of our set with reddish colour, probably due to iron oxide generated by weathering in the Lybian desert. \\
Allende is classified as CV\textsubscript{OX}. This group displays Calcium-Aluminum-rich Inclusions (CAIs) and ameboid olivine refractory elements and this suggests they have been formed at high temperature during the early stages of Solar System formation. A recent discovery of a class of asteroids (the so-called Barbarians) seems to have a composition unusually enriched in CAI minerals, like spinel \citep{cellino2014}.
Allende differs from other CV\textsubscript{OX} meteorites in its lower matrix-chondrule ratio and in having chondrules with more opaque minerals \citep{brearley}. 

\paragraph*{FRO95002 and FRO99040}
These two meteorites were found in Antarctica, in the region of the Frontier Mountain in 1995 and 1999, respectively.  Both belong to the CO3 class, which presents similar characteristic of CV3 regarding chondrules and matrix abundance, but differ in the chondrule dimensions, which in the CO3 case are $\mu$m-sized. 
The CO chondrites are all of petrologic type 3, showing slightly different metamorphic stages, ranging from 3.0 to 3.7 \citep{mcsween1}. 
CO chondrites usually have low Fe compositions with olivine with Fa<1 \citep{brearley}. 
Also Fe-pyroxenes are rare, only a few pyroxenes with Fs$\sim$10 are reported \citep{brearley}. 

\subsubsection{Minerals}

The first hint of the presence of silicates in the cometary dust dates back to the observation of the emission feature at 10 $\mu$m in comet Bennett and its consequent laboratory modeling \citep{maas_bennett}. The measurements of 1P/Halley mineralogical composition revealed Mg-rich olivine (forsterite) and pyroxene (enstatite) as dominant compounds from in-situ mass spectra of particles provided by Giotto and Vega Spacecraft \citep{schulze_halley}. Crystalline forsterite and enstatite have been detected also on comet Hale-Bopp  \citep{crovisier_halebopp} and   9P/Tempel from space-based infrared spectra \citep{lisse_tempel} and on 81P/Wild2 from direct measurement of particles collected by Stardust mission  \citep{zolensky_wild2006}. These silicates are predicted by thermodynamic models to condense in a hot gas at 1200-1400 K, while the reaction with Fe occurs only at lower temperature. Thus, the preponderance of Mg-rich silicates in cometary dust can be explained by direct condensation in the inner primordial nebula \citep{hanner}. 
The olivine and pyroxene samples used in this work have respectively the composition Mg\textsubscript{1.85}Fe\textsubscript{0.14}SiO\textsubscript{4} and Mg\textsubscript{0.85}Fe\textsubscript{0.08}Si\textsubscript{0.99}O\textsubscript{3}, and are therefore close to the Mg-rich endmembers, i.e. forsterite and enstatite. Thus, in the following we will refer to the olivine and pyroxene samples as forsterite and enstatite. The forsterite sample is the one identified as 'Olivine S' in \cite{munoz_allende}. \\

\subsubsection{Organics}
Tholins are organic compounds generated by irradiation (solar UV, high energy particles, electrons)  of mixtures of common compounds of carbon, oxygen and nitrogen, e.g. CO$_2$, CH$_4$ and HCN \citep{sagan}.
They are thought to be present on the surfaces and/or in the  atmospheres of several Solar System bodies. For example, they are
likely responsible for the reddish color of many objects such as Pluto \citep{gladstone_tholins_pluto} and Ceres \citep{combe_tholins_ceres}, and are believed to be responsible for Titan's atmospheric haze \citep{brasse_tholins_titan}.\\
Analysis of 1P/Halley dust mass spectra indicated for the first time the existence of CHON particles in comets \citep{kissel_halley}. The complex nature and variety of organics in cometary dust was revealed by analysis of samples returned from 81P/Wild, obtained by \textit{Stardust} \citep{clemett_stardust2010, sandford2006, matrajt_sturdust2008}. More recently, the in situ analysis of 67P coma dust by COSIMA/\textit{Rosetta} mass spectrometer has confirmed the presence of abundant high-molecular weight organic matter, nearly 50\% in mass \citep{bardyn_cosima67P, fray2017}.
Thus, the abundance of CHON compounds in cometary dust suggests that tholins are likely to form in this environment as well.
Tholins have been previously used in laboratory studies as cometary dust simulants, often mixed with other components, typically silicates and ices \citep{poch_tholins_ice,poch_tholins_silicates,jost_spectra67P}.\\
The tholins sample used in this work was synthesized using the PLASMA experimental setup described previously by \cite{brasse2017}. It allows the production and sampling of tholins from N$_2$-CH$_4$ electron irradiation in a glove box without contamination by the air of the laboratory. 
The elemental composition of these tholins has been determined  \citep{brasse_thesis}. They are carbon rich (C/N = 2.2 to 2.4) with a C/H of 0.7 to 0.8. Several grams of tholins were prepared to perform the CODULAB measurements.


\subsection{Sample preparation}\label{sample preparation}

In order to obtain particulate samples with different size distributions within the typical size range of cosmic dust particles, the DaG521, FRO95002 and FRO99040 samples were milled and dry-sieved at the Department of Geoscience of the University of Padua. After milling, the three samples were first size-segregated using a 71 $\mu$m sieve. The fraction of sample that passed through the sieve was subsequently sieved using a 53 $\mu$m sieve. This procedure generated three subsamples designated as DaG521M, FRO99040M and FRO95002M with diameters, $d$, in the range 53 $\mu$m < $d$ < 71 $\mu$m. The fractions of DaG521M and FRO99040M that passed through the 53 $\mu$m sieve were subsequently sieved through a 45 $\mu$m sieve producing subsamples DaG521S and FRO99040S (45 $\mu$m < $d$ < 53 $\mu$m). 
 Not enough material of FRO99040M was left to produce a further sample.
A similar sieving procedure was used to produce the olivine samples described by \cite{munoz_allende}. 

As far as the organic sample is concerned, after their synthesis, the tholins were kept in gas proof vials under dry N$_2$ atmosphere, to avoid their chemical evolution. Just before measurements with CODULAB the tholins were first homogenized with a mortar to break the agglomerates and sieved to obtain an homogeneous powder. Then, these tholins were introduced into the cylindrical stainless steel tank of the aerosol generator.

\section{Results and discussion}

\subsection{Sample characterization}

\begin{table}
	\centering
	\begin{tabular}{ccc}
		Sample & Refractive Index & Appearance \\
		\hline \hline
		DaG521 & $1.65 + i10^{-3} $ & reddish\\
		\hline
		FRO99040 & $1.65 + i10^{-3} $ & dark grey \\
		\hline 
		FRO95002 & $1.65 + i10^{-3} $ & dark grey \\
		\hline
		Allende & $1.65 + i10^{-3} $ & dark grey \\
		\hline 
		Forsterite & $1.62 + i10^{-5} $ & white\\
		\hline
		Enstatite &  $1.58 + i10^{-5} $ & white\\
		\hline
		Tholins & $1.35 + i2.3\cdot10^{-2}$ & dark \\
		\hline\hline
	\end{tabular}
	\caption{Refractive index of each sample and their relative appearance.}
	\label{tab:ref_index}
\end{table}
\begin{figure}
    \centering
    \includegraphics[width =0.48
    \textwidth]{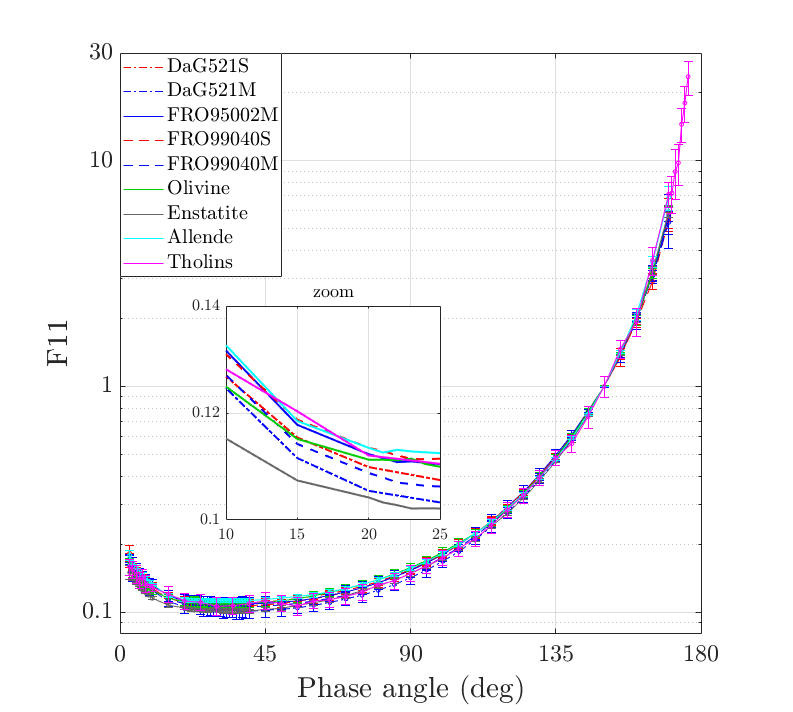}
    \caption{Phase functions of all samples at 520 nm. The inset shows a zoom-in of the backscattering region.}
    \label{fig:phase_comparison_all}
\end{figure}

\begin{figure}
    \centering
    \includegraphics[width =0.48
    \textwidth]{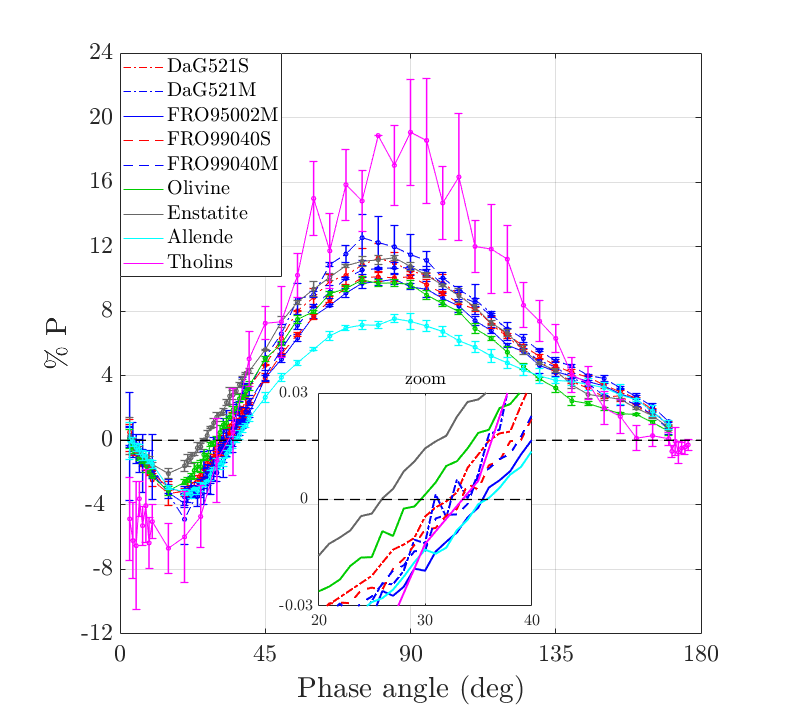}
    \caption{Polarization curve of all samples at 520 nm. The inset shows a zoom-in of the inversion angle region.}
    \label{fig:tot_comparison}
\end{figure}
\begin{figure*}
    \centering
    \includegraphics[width = \textwidth]{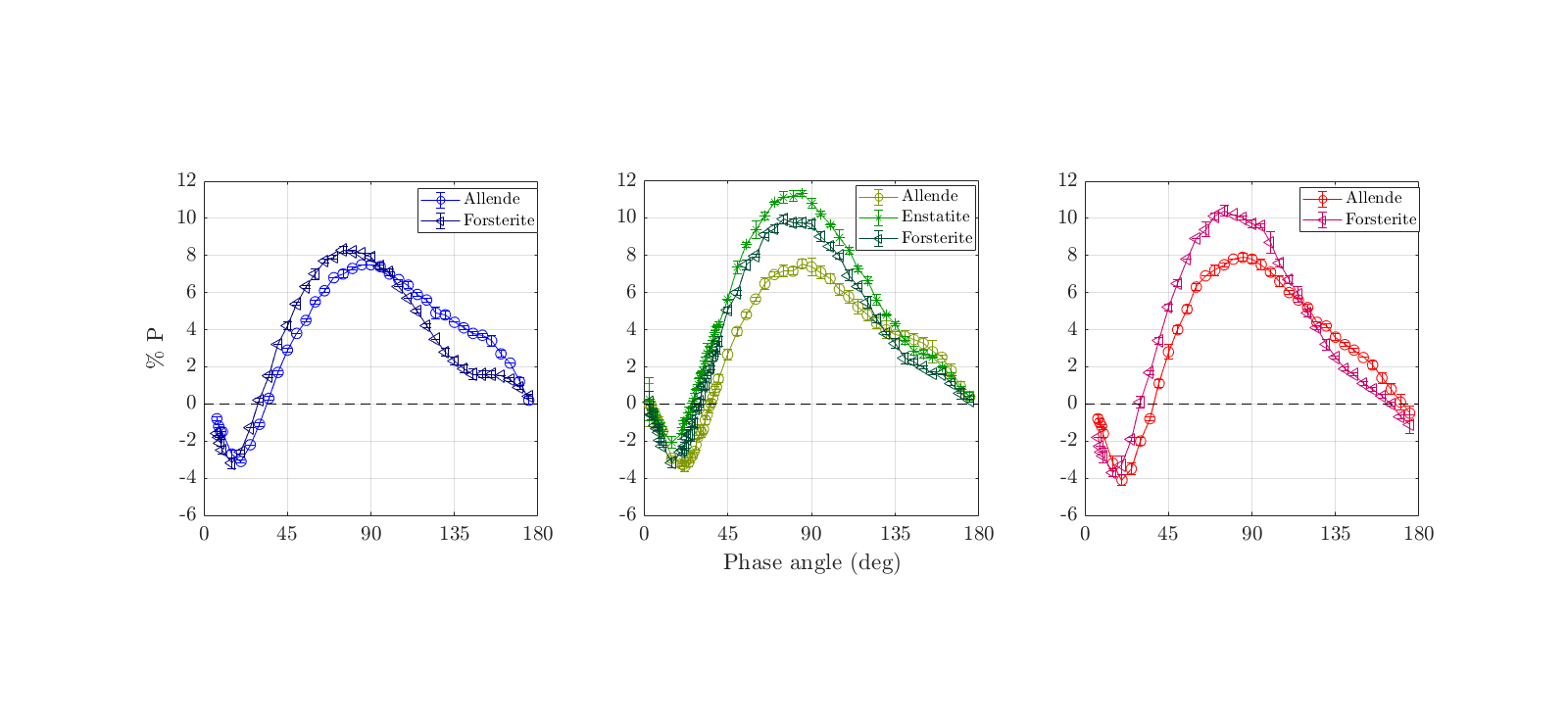}
    \caption{Measured degree of linear polarization curves for Allende meteorite (circles), forsterite (triangles), and enstatite (stars) at three different wavelengths 442 nm (left panel), 520 nm (center panel), and 633 nm (right panel).}
    \label{fig:tris}
\end{figure*}

\subsubsection{Refractive index}\label{refractive_index}
The refractive index of a medium is defined as:
\begin{equation}
m = c\sqrt{\epsilon \mu} = n + ik
\label{eq:index}
\end{equation}
where $\epsilon$ is the electric permittivity, $\mu$ is the magnetic permeability and $c$ the speed of light in vacuum. The optical constants $n$ and $k$ represent, respectively, the phase velocity of the wave in the medium and the absorption coefficient of the material. \\
Estimates of the refractive indices of our samples are compiled in Table \ref{tab:ref_index}. The refractive indices of forsterite and enstatite (\cite{m_silicates}) have been obtained from the Jena-St. Petersburg Database of Optical Constants (http://www.astro.uni-jena.de/Laboratory/Database/jpdoc/index.html).
The refractive index of tholins at 532 nm has been reported by \cite{hasenkopf_index_tholin}; for a detailed analysis of the tholins optical properties see \cite{brasse_tholins_titan}. \\
The refractive indexes of the Allende, DaG521, FRO99040, and FRO95002 meteorite samples are unknown. Therefore,  we have assumed estimates based on literature values of the main constituents. The imaginary part of the refractive index of the terrestrial samples is significantly smaller than that of the meteorites and tholins, which indicates a smaller absorbance of these materials, consistent with their light colour. By contrast, the meteorite powders are darker as a result of the higher bulk Fe content  and organic components. \\
It should be kept in mind that the refractive index generally depends on the wavelength of the incident light. While the Mg-rich forsterite and enstatite present a flat wavelength dependence in the visible, the Fe content of the meteorite samples is expected to produce an effect on the refractive index at different visible wavelengths.
This fact has been recently taken into account by \cite{devogele} to interpret the observed wavelength dependence of the inversion angle of polarization of asteroid (234) Barbara as a consequence of the imaginary
refractive index, in turn a consequence of the presence of nano-iron
phase particles.

\begin{table*}
	\centering
	\begin{tabular}{ccccccccc}
		Sample & $r_{eff} \left(\mu{\rm m}\right)$  & $v_{eff}$ & $P_{min}(\%)$ & $\alpha_{min}(deg)$ &  $\alpha_{0}(deg)$ & $P_{max}(\%)$ & $\alpha_{max}(deg)$ & h(\%/deg)\\
		\hline \hline
		\textbf{DaG521S}   & 3.58 & 1.96  &$-3.3\pm0.7$& $15\pm5$ & $32\pm1$ & $11.3\pm0.91$ & $80\pm5$& 0.24  \\
		\hline 
		\textbf{DaG521M}  & 8.69 & 2.43  &$-4.9\pm1.6$ & $20\pm5$ & $34\pm1$ & $12.6\pm1.4$ & $75\pm5$& 0.31  \\
		\hline 
		\textbf{FRO95002M} & 3.92 & 2.72 &$-3.9\pm0.1$ & $20\pm5$ & $35\pm1$ & $10.0\pm0.3$ & $85\pm5$ & 0.20  \\
		\hline
		\textbf{FRO99040S}& 3.68 & 1.72  &$-3.3\pm0.3$ & $20\pm5$ & $33\pm1$ & $10.1\pm0.3$ & $80\pm5$ & 0.21  \\
		\hline
		\textbf{FRO99040M} & 5.90 & 3.10  &$-3.5\pm0.6$ & $20\pm5$ & $34\pm1$ & $10.7\pm0.1$ & $90\pm5$ & 0.19  \\
		\hline 
		\textbf{Forsterite}& 3.06 & 1.04  &$-3.1\pm0.3$ & $15\pm5$ & $30\pm1$ &  $9.9\pm0.2$ & $75\pm5$ & 0.22  \\
		\hline
		\textbf{Enstatite}& 3.70 & 3.13  &$-2.1\pm0.3$ & $15\pm5$ & $26\pm1$ & $11.3\pm0.2$ & $85\pm5$ & 0.19 \\  
		\hline
		\textbf{Allende}  & 2.44 & 3.42 &$-3.3\pm0.3$ & $22\pm1$ & $36\pm1$ &  $7.5\pm0.2$ & $85\pm5$ & 0.15\\ 
		\hline 
		\textbf{Tholins}  & - & - & $-6.7\pm1.5$ & $15\pm5$ & $35\pm1$ & $19.1\pm3.3$ & $90\pm5$ & 0.35 \\
		\hline\hline
	\end{tabular}
	\caption{Polarimetric parameters of the samples. 
		$r_{eff}$  and $v_{eff}$ are the effective radius and effective variance. $P_{min} $  is the minimum of polarization  and $\alpha_{min} $ the relative phase angle.   $P_{max} $  is the maximum of polarization at phase angle $\alpha_{max} $ and $\alpha_{0}$ is the inversion angle.  h(\%/deg) is the slope of the polarization curves computed between  $\alpha_{max}$ and $\alpha_0$. Tholins measurements were taken in the red domain, at $\lambda = 632 $nm. }
	\label{tab:table_pol}
\end{table*}

\subsubsection{Size distribution}
 The size distributions of our samples have been estimated using a low angle laser light scattering (LALLS) particle sizer (Malvern Mastersizer 2000, \cite{malvern}). The LALLS method relies on the measurement of the phase function of samples dispersed in a carrier fluid at 633 nm within a range of low scattering angles (0.02$^{\circ}$-30$^{\circ}$). The volume distribution of equivalent spherical particles that best reproduces the observed phase function is obtained by inverting a light scattering model based on Mie theory, which requires knowing the complex refractive index of the samples (eq. \ref{eq:index}). Since the refractive indexes of our meteorite samples are unknown, we have carried out a sensitivity study of the impact on the retrieved size distributions of varying $n$ and $k$ within wide ranges. In these sensitivity tests we have used as reference value the estimated refractive index for the Allende meteorite ($m=1.65, k=i0.001$) \citep{munoz_allende}. First, the imaginary part of the refractive index, $k$, was fixed to the reference value ($k=0.001$). The value of real part, $n$, was then varied between 1.5 and 1.7 in steps of 0.05. Second, the real part was fixed to the reference value ($n=1.65$), and the imaginary part, $k$, was varied from $10^{-5}$ to $10^{-1}$ in factor of 10 steps.
 Artifacts are found  in the retrieved size distributions in the small size range  (r < 1 $\mu$m)  when low values of the real  part ($n=1.5$; $n=1.55$) or extreme values of the  imaginary part ($k=10^{-5}$; $k=10^{-2}$; $k=10^{-1}$) are assumed. Therefore, we consider the reference value $n=1.65+i0.001$ as a reasonable estimate of the refractive index of our meteorite samples.\\
In spite of the sieving procedure described in Section \ref{sample preparation}, the meteorite samples show  broad size distributions  with volume equivalent radii ranging from 0.3 $\mu m$ to $\sim$ 100 $\mu m$. Figs. \ref{fig:sd_minerals} and \ref{fig:sd_meteorites_2} show the $S(logr)$ size distributions of the samples. Here, $S(logr)$ is the projected-surface-area for a volume equivalent sphere with radius $r$. Size distributions are commonly characterized by the effective radius $r_{eff}$ and effective variance $v_{eff}$ as defined by \cite{hansen}.
 These parameters have a straightforward interpretation for mono-modal distributions, while for multi-modal distributions, the $r_{eff}$ and $v_ {eff}$ are only first order indicators of the particles size. For example, the DaG521M and FRO99040M samples show bi-modal size distributions with secondary peaks at $\sim$ 30 $\mu m$ and $\sim$ 20 $\mu m$, respectively. \\
Due to the limited amount of the tholins sample, we could not measure its size distribution. Based on the SEM images  (Figs. \ref{fig:sem_images}k and \ref{fig:sem_images}l) we can estimate a broad size distribution with sizes ranging from sub-micron up to hundred of microns.

\subsubsection{Morphology} 
Fig. \ref{fig:sem_images} shows Field Emission Scanning Electron Microscope (FESEM) images of our samples. It can be seen that in all cases the particles present a wide variety of irregular shapes. Figs. \ref{fig:sem_images}a - \ref{fig:sem_images}d show that the enstatite, forsterite and Allende meteorite particles present sharp edges and relatively clean flat surfaces compared with the other samples. The DaG521, FRO95002 and FRO99040 particles (Figs. \ref{fig:sem_images}e - \ref{fig:sem_images}j) present more rounded shapes with rough surfaces covered by a layer of small particles. These morphological differences could be partially caused by the powder preparation method. The dry-sieving procedure could not remove small particles that remained stuck on the surface of the large particles by electrostatic forces. By contrast, the enstatite, forsterite and Allende meteorite samples underwent wet-sieving, which removed a large fraction of the finest particles from the original sample. For this reason the grain surfaces appear generally smoother than for the other samples. Tholins show peculiar structures due to the synthesis process that produced very elongated particles of the order of 10 $\mu m$ with layered substructure and sub-$\mu m$ surface roughness (Figs. \ref{fig:sem_images}k - \ref{fig:sem_images}l).

\subsection{Measurements}\label{measurements}


In this section we present the measured phase functions $F_{11}(\alpha)$ (Fig. \ref{fig:phase_comparison_all}) and degree of linear polarization for unpolarized incident light $P(\alpha)=-F_{12}(\alpha)/F_{11}(\alpha)$ (Fig. \ref{fig:tot_comparison}) for all samples studied in this work. 
The measurements were performed at a wavelength of 520 nm covering the phase angle range from 3$^{\circ}$ to 175$^{\circ}$. As mentioned, the measured phase functions are normalized to unity at $\alpha=150^{\circ}$. The experimental phase function and polarization curves are freely available in the Amsterdam-Granada Light Scattering Database (www.iaa.es/scattering) under request of citation of this paper and \cite{munoz_database}.

The light scattering behavior of the samples considered in this work is qualitatively similar to that of other types of irregular mineral dust investigated in the laboratory (see \cite{munoz_database} and references therein). The measured phase functions show a strong peak at large phase angles, a plateau with almost no structure at intermediate angles and a  moderate increase at small phase angles. Fig. \ref{fig:phase_comparison_all} does not reveal large differences between samples.

The measured $P$ curves display the typical bell-shape with a small negative branch starting at an inversion angle, $\alpha_0$ and reaching a minimum at $\alpha_{min}$ (Fig. \ref{fig:tot_comparison}). The wide positive branch has a maximum $P_{max}$ at $\alpha_{max} = 75^{\circ}-90^{\circ}$.
These polarization curves are also similar to those observed by remote sensing for a variety of Solar System objects such as comets and asteroids. In contrast to the phase functions, the polarization curves of our samples do show significant variability. Table \ref{tab:table_pol} summarizes the main parameters of the degree of linear polarization curve in the region of minimum polarization ($P_{min}, \alpha_{min}$), inversion angle ($\alpha_0$), and maximum polarization ($P_{max},\alpha_{max}$). \\

Among our samples, tholins show the highest maximum values of linear polarization ($P_{max} = 19\%$) and the deepest negative branch ($P_{min} = -6.7\%$). Note that the empirical relationship known as \textit{Umov's effect} indicates an inverse relationship between the maximum value of polarization and the geometric albedo \citep{zubko2011}, and tholins are indeed the darkest material considered in this study. However, for a cloud of irregular particles in single scattering conditions the interrelation between $P_{max}$ and the geometric albedo is dependent on the size distribution of the particle cloud \citep{zubko_umov2017,zubko_umov2018}. Thus, the Allende meteorite sample, which has the smallest effective radii, also presents the lowest maximum of the degree of linear polarization in spite of its dark color. Interestingly, the polarization curve of the Allende sample has a distinct feature compared to the pure silicate samples, showing a shoulder near $\alpha = 150^{\circ}$  at all wavelengths (see Fig. \ref{fig:tris}).  The measured values of $\alpha_0$ range from 26$^{\circ}$ for enstatite to 36$^{\circ}$ for the Allende sample. We do not find a clear relation between size distribution and  position of the inversion angle. By contrast, for the range of sizes of our samples, the location of the inversion angle seems to be dependent on the composition. Terrestrial materials consisting of pure non-absorbing minerals show lower inversion angles than the meteorites and tholins, which present iron and organic material in different proportions.

A fully satisfactory explanation of the negative branch of the linear polarization curve is still missing. Numerous Solar System objects such as cometary comae or asteroids and satellites, which are covered by a thick regolith layer, show this feature. 
Even quite small asteroids, for which we might assume that the surface regolith layer is thinner, generally exhibit the same general polarimetric behaviour as larger asteroids.\\
In principle, each powdered material should produce negative values of polarization at low phase angles, when coherent backscattering by small particles is considered. 
This physical effect results from  the interference of wavelets scattered by the particulate medium. 
At zero phase angle the wavelets are in phase and they combined positively enhancing the intensity. 
The linear polarization reflects this effect as well, inverting the plane of wave polarization and favoring the parallel component of electromagnetic wave to be scattered, rather than the perpendicular one. The role  of coherent backscattering in the development of a negative polarization branch and a backscattering enhancement in the phase function has been discussed by \cite{muinonen_backscattering} using the exact method known as the  superposition T-Matrix \citep{cb}.\\

The origin of the negative branch of clouds of particles in single scattering conditions and corresponding particulate surfaces has also been investigated in several laboratory studies (e.g. \cite{shkuratov2006,shkuratov2007,jesus_lunar}). In a recent work,  \cite{jesus_lunar} compare the scattering behavior of two samples of the same lunar simulant  with different size distributions.
The effect of removing particles smaller than $\sim 1$ $\mu m$ from the pristine  sample on the polarization curve is noticeable. The maximum of the  linear polarization increases by a factor of 1.5. 
Apparently the small particle fraction of the pristine sample was limiting the maximum of the degree of linear polarization. 
Further, the negative polarization branch (absolute values) is decreased from 2.4\%  to 0.8\% after removing particles smaller than 1 $\mu$m. This result seem to indicate that sub-micron  scale features might be responsible for the negative  branch of the degree of linear polarization. \\
In summary, it is quite difficult to disentangle all the effects responsible for the polarization curve shape, we can only highlight some empiric relations, but it is essential to further investigate the subject.
In sections \ref{BSE} and \ref{size_section} the differences observed in the phase function and degree of linear polarization are discussed in terms of different physical properties (size, color, shape), in order to investigate their influence on the scattering behavior of the samples. 

\begin{table*}
\centering
\begin{tabular}{ccccccc}
Sample & $r_{eff} \left(\mu{\rm m}\right)$  & $v_{eff}$ &  $x_{eff}$ & $F_{11}^{syn}(0^{\circ})$ & $F_{11}^{syn}(30^{\circ})$ &  BSE (520 nm) \\
\hline \hline 
\textbf{DaG521S} &  3.58 & 1.96  & 43.3 & 0.384 & 0.191 & 2.01\\
\hline
\textbf{DaG521M} &   8.69 & 2.43  & 105.0 &  0.419 & 0.214 & 1.96 \\
\hline 
\textbf{FRO95002M}&   3.92 & 2.72 & 47.4 & 0.390 & 0.201 & 1.94\\
\hline
\textbf{FRO99040S} & 3.68 & 1.72  &44.5 & 0.612& 0.270 & 2.27 \\
\hline 
\textbf{FRO99040M}   & 5.90 & 3.10  &71.3 & 0.427 & 0.203 & 2.10\\
\hline 
 \textbf{Forsterite} & 3.06 & 1.04  & 37.0 & 0.384 & 0.197 & 1.95 \\
\hline 
 \textbf{Enstatite} & 3.70 & 3.13  & 44.7 & 0.369 &0.188 & 1.96\\
\hline
 \textbf{Allende}   & 2.44 & 3.42 & 29.5 &  0.519 & 0.266 & 1.95\\
\hline\hline
\end{tabular}
\caption{Parameters related to the backscattering enhancement computed for the samples at different wavelengths. BSE= $F_{11}(0^{\circ})/F_{11}(30^{\circ})$.}
\label{tab:table_phase}
\end{table*}

\begin{figure} 
    \centering
    \includegraphics[width=0.48\textwidth]{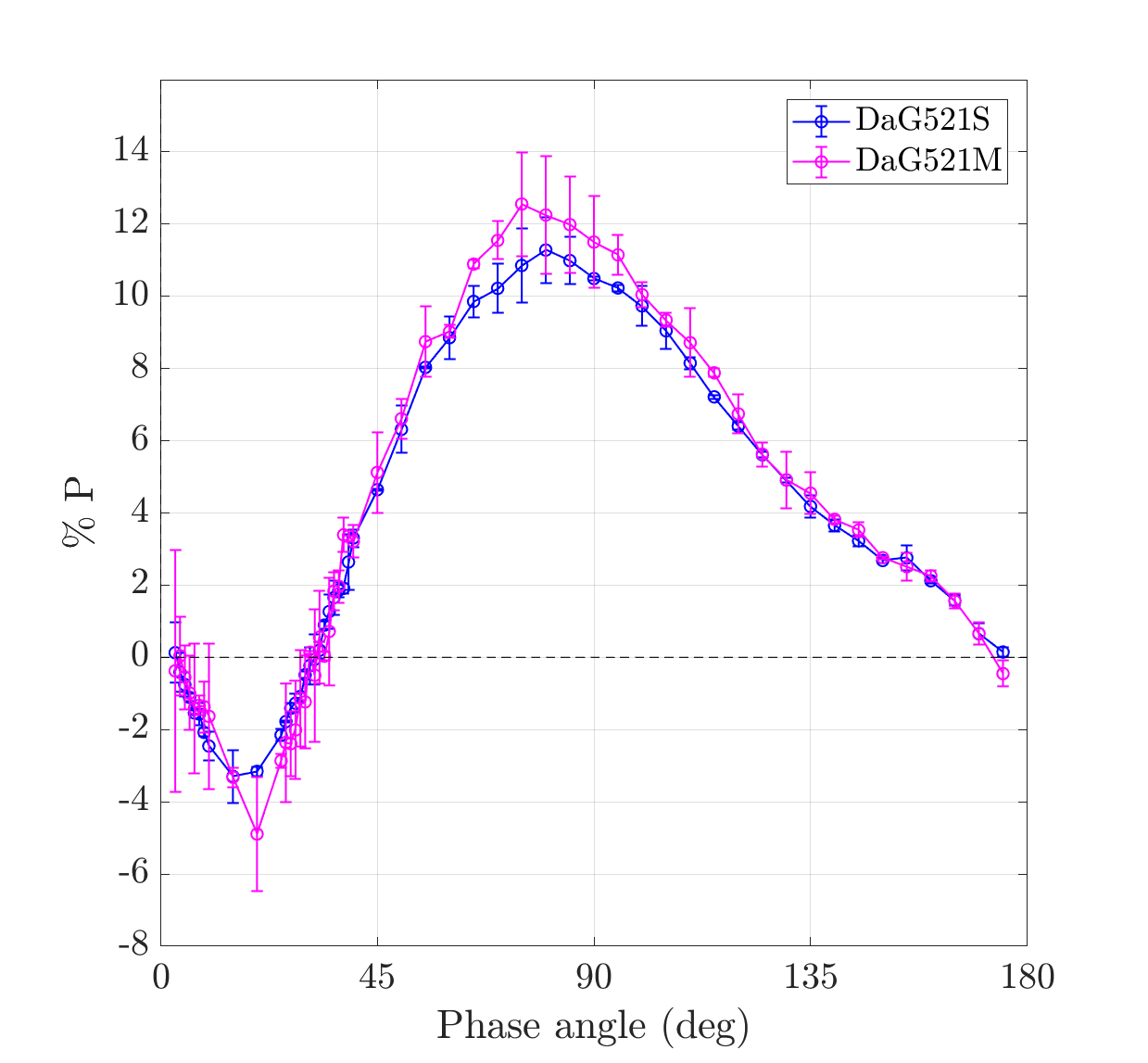}
    \caption{Measured degree of linear polarization curves for the DaG521S (blue symbols) and DaG521M (magenta symbols) samples at 520 nm. }
    \label{fig:dag_comparison}
\end{figure}


\begin{figure} 
    \centering
    \includegraphics[width =0.48 \textwidth]{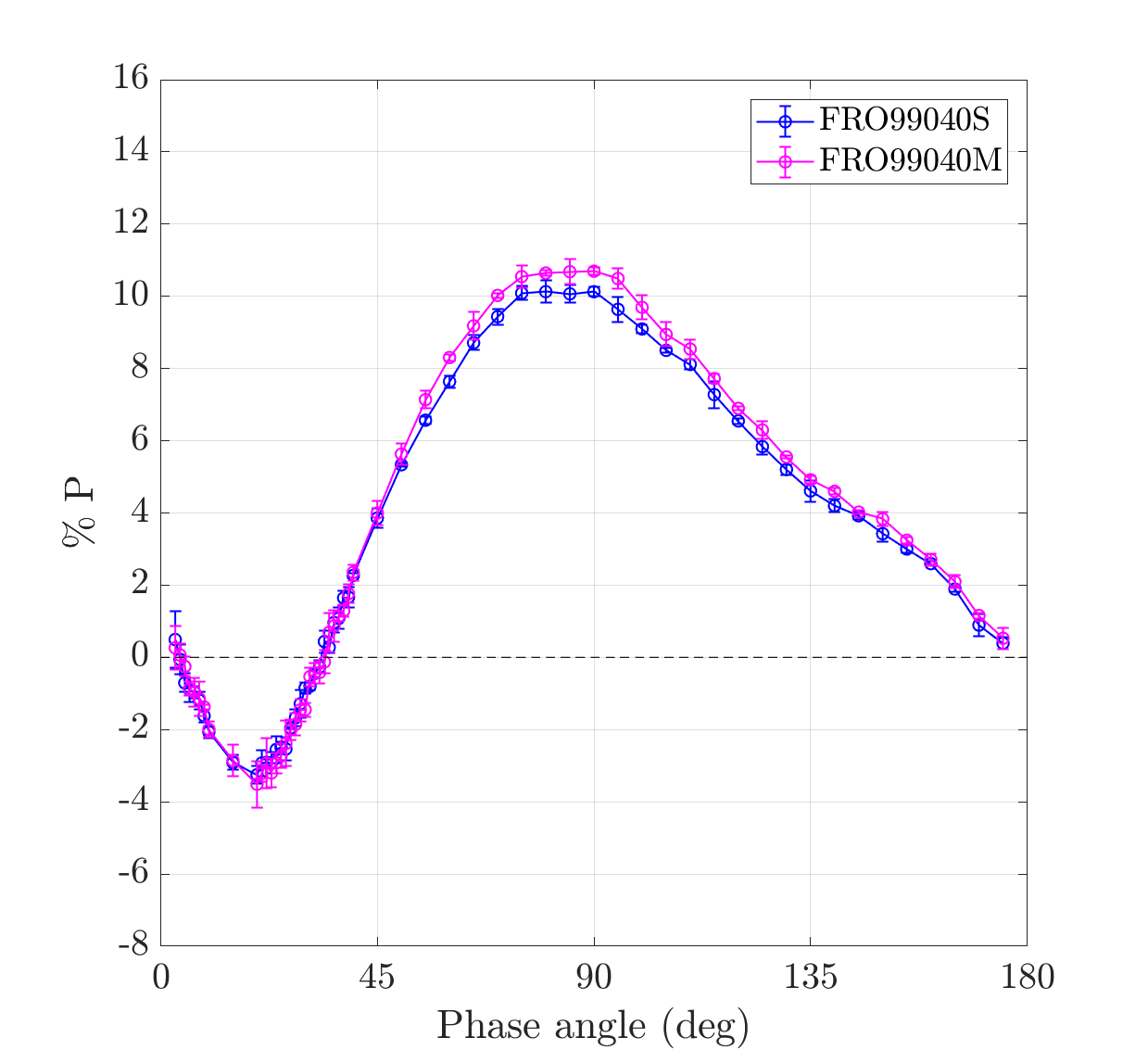}
    \caption{Measured degree of linear polarization curves for the FRO99040S (blue symbols) and FRO99040M (magenta symbols) at 520 nm.}
    \label{fig:fro99_comparison}
\end{figure}

\subsection{Backscattering enhancement}
\label{BSE}
In order to give an estimation of the trend of the phase function in the backward  region  for all studied samples, we compute the backscattering enhancement (BSE), defined as the ratio of measured phase function values at $\alpha = 0^{\circ}$ and  $\alpha = 30^{\circ}$ \citep{bertini_2007}. Since our laboratory measurements do not cover the whole phase angle range we extrapolated a synthetic phase function, 
$F_{11}^{syn}$, from our measurements that range from $3^{\circ}$ to $175^{\circ}$. The synthetic phase function is defined in the full angle range from $0^{\circ}$ to $180^{\circ}$ and is normalized according to: 




\begin{equation}
\label{normaliza}
\frac{1}{2}\int_{0}^{\pi} d\alpha sin\alpha F_{11}^{syn}(\alpha)=1. 
\end{equation}

The extrapolation of the measured phase function in the forward direction is based on the assumption that the forward scattering peak for randomly oriented particles with moderate aspect ratios is mainly dependent on the size and refractive indices of the particles, but not on their shapes \citep{liu2003}. Between $175^{\circ}$ and $180^{\circ}$, we produce Mie calculations for projected-surface-area equivalent spheres. For the Mie computations we use the measured size distribution of the corresponding sample and  the value of the refractive indices presented in Tab. \ref{tab:ref_index}. For the backward direction, we first generate a value for the phase function at $0^{\circ}$ by a quadratic function generated by least squares  with the measured data points from $3^{\circ}$ to $30^{\circ}$. Then all values between $3^{\circ}$ and $30^{\circ}$ are produced by a cubic splines interpolation considering an additional condition that must be fulfilled in all cases: the first derivative of the phase function at $0^{\circ}$ must be zero \citep{hovenierGuirado2014}. At this stage, both the forward peak ($175^{\circ}-180^{\circ}$) and the rest of the phase function ($0^{\circ}-175^{\circ}$) are defined, but they are normalized in a different way: the forward peak belongs to a function normalized according to eq. \ref{normaliza} and the normalization of the rest of the function is arbitrary. The function defined by the measured ($3^{\circ}-175^{\circ}$) plus the extrapolated ($0^{\circ}-3^{\circ}$) data points is then vertically shifted  until the computed $F_{11}^{syn}(175^{\circ})$  matches the measured $F_{11}(175^{\circ})$. The normalization condition (eq. \ref{normaliza}) is then checked. If it is not satisfied within a 0.1$\%$ accuracy, the value of the measurement at $175^{\circ}$ is increased or decreased (within the experimental error bars) until the normalization condition is fulfilled. The backscattering enhancement for all our samples lays in the range [1.94-2.27]  (see Table  \ref{tab:table_phase}).

\subsection{The effect of size and color on polarization} \label{size_section}
The degree of linear polarization of the DaG521S and DaG521M, and the FRO99040S and FRO99040M samples are plotted  in Figs. \ref{fig:dag_comparison} and \ref{fig:fro99_comparison}, respectively.    
Since each figure refers to the same material, the differences observed are only attributable to the different size distributions of the S and M samples. The salient feature of this comparison is an increase of the maximum polarization with size. In fact, it can be seen in Fig. \ref{fig:sd_meteorites_2} that the size distributions of the samples with a higher $P_{max}$ (DaG521M  and FRO99040M) present a secondary maximum at larger sizes. Thus, the contribution of particles with large size parameters ($x_{eff}=2\pi r_{eff}/\lambda$) is most likely responsible for the measured increase in $P_{max}$.\\ 
The direct relationship between $P_{max}$ and $r_{eff}$ (or $x_{eff}$) apparent in Figs. \ref{fig:dag_comparison} and \ref{fig:fro99_comparison} suggests an inverse relationship between  $P_{max}$ and wavelength, since the size parameter is inversely proportional to wavelength. \\

The polarization curves for the forsterite sample at three different wavelengths (442 nm, 520 nm, and 633 nm) are plotted in Fig. \ref{fig:oli_col}. The measurements at 442 nm and 633 nm have been published  previously \citep{munoz_allende}. It can be seen that $P_{max}$ increases as the size parameter of the particles decreases ($\lambda$ increases for a fixed $r_{eff}$). In this case the differences in the measured degree of linear polarization are also likely related to the size parameter of the grains, since the refractive index of forsterite presents a flat dependence on wavelength in the studied region. \\
The apparent contradiction between Figs. \ref{fig:dag_comparison} and \ref{fig:fro99_comparison} on one side and  Fig. \ref{fig:oli_col} on the other can be explained by the broad size range of our samples. The scattering properties of dust grains are described by three regimes: Rayleigh, resonance and geometric optics. When particles are smaller than the wavelength (Rayleigh regime), $P_{max}$ tends to decrease as the size parameter increases. In that case, $P_{max}$ increases with wavelength if the refractive index is constant, i.e.  $P_{max}\propto 1/x \propto\lambda$. 
By contrast, in the geometric optics regime ($x_{eff}\gg \lambda$), $P_{max}$ tends to increase as the size parameter becomes larger, i.e. $P_{max}\propto x \propto 1/\lambda$. These trends are well illustrated by computed polarization data for Gaussian random shapes from the Rayleigh to the geometric optics regimes reported by \cite{liu_gaussian_sphere}. As shown  in Fig. \ref{fig:sd_meteorites_2}, the contribution of grains in the geometric optics domain for samples DaG521M and FRO95002M is relatively large. 
Therefore, the $P_{max}$ tends to increase as $x_{eff}$ increases (Figures \ref{fig:dag_comparison} and \ref{fig:fro99_comparison}). 
The majority of the grains in the forsterite sample belong to the Rayleigh and resonance ($x_{eff}\approx\lambda$) domains where the dependence of $P_{max}$ with size is nearly opposite, i.e. it increases as $x_{eff}$ decreases (Fig.\ref{fig:oli_col}).

The degree of linear polarization curves for the Allende meteorite sample at three wavelengths (442 nm, 520 nm, and 633 nm) are plotted in Fig. \ref{fig:all_col}. As shown In Fig. \ref{fig:sd_minerals}, the Allende sample mainly consists of particles in the Rayleigh and resonance regimes. In this case there is no clear trend of $P_{max}$ with wavelength.
The higher Fe content of the Allende meteorite compared to forsterite indicates a higher imaginary part of the refractive index at shorter wavelengths, implying a higher reflectance at red wavelengths. 
According to the Umov effect, the larger reflectance, the lower the maximum of the degree of linear polarization. 
Thus, the increase of $P_{max}$  with wavelength seems to be balanced by a lower albedo at shorter wavelengths. 
Therefore, the maximum of the degree of linear polarization is dependent not only on the size of the grains but also on their refractive index.
 A study of tholins generated in plasma with different size distributions by \cite{hadamcik2009} is in line with our results, showing that in Rayleigh regime $P_{max}$ decreases when the particle size increases, while in optical regime $P_{max}$ increases with the particle size.\\
It is worth noting that in the case of the forsterite sample (Fig. \ref{fig:oli_col}), the inversion angle $\alpha_0$  does not change with wavelength. However, some wavelength dependence of the negative branch minimum can be distinguished at 633 nm. 
In the case of the Allende sample, both the minimum of the negative branch and the inversion angle vary with wavelength. The apparent dependence of the negative branch parameters on composition (i.e. refractive index) seems to be in agreement with the simulations of \cite{zubko_absorption,zubko_index}, which demonstrate that the negative polarization parameters strongly depend not only on size parameter but also on the material absorption properties.

\begin{table*}
\centering
\begin{tabular}{cccccccc}
\textbf{Comet} &$\lambda$ (nm) &  $P_{min}(\%)$ & $\alpha_{min}(^\circ)$ & $\alpha_0(^\circ)$  &  $P_{max}(\%)$ & $\alpha_{max}(^\circ)$ & h ($\%$/deg)\\
\hline \hline 
High $P_{max}$ & 515 & $-1.5\pm0.5$ & $9\pm2$ & $22.2\pm0.5$ & $26\pm2$ & $103\pm10$& $0.22\pm0.02$ \\ 
& 670 & $-1.5\pm0.5$ & $11\pm2$ & $22.6\pm0.5$ & $28\pm3$ & $95\pm10$& $0.25\pm0.03$\\
\hline  
Low $P_{max}$ &515&$-1.7\pm0.5$ & $6\pm3$ & $19.0\pm0.5$ & $10\pm3$ & $80\pm10$& $0.20\pm0.02$ \\
 &670&$-1.9\pm0.5$ & $6\pm3$ & $20.5\pm0.5$ & $18\pm3$ & $95\pm10$& $0.22\pm0.02$ \\
 \hline  
67P  &red & $-1.7\pm0.1$  & $12\pm3$ & $22\pm2$ &- & -& $0.35\pm0.02$* \\ 
\hline\hline
\end{tabular}
\caption{Polarimetric parameters of two classes of comets from \protect\cite{levasseur-regourd_class}. Values for 67P correspond to post-perihelion period and are taken from  \protect\cite{hadamcik}. *The $h$ value for 67P is the slope at the inversion angle, since no value for  $P_{max}$ are available.}
\label{tab:table_comet}
\end{table*}

\subsection{Comparison with asteroids and comets}


  Ground-based observations of the intensity of light scattered by cometary dust grains are usually limited to certain observational geometries. Moreover, time variations of the brightness of the coma as observed from Earth do not only depend on the phase angle but also on changes of the dust production rate as the comet moves in its orbit around the Sun (see e.g. \cite{kolokolova_comets2}). The OSIRIS camera on board the Rosetta spacecraft revealed the behavior of comet 67P/Churyumov-Gerasimenko phase functions from inside the coma \citep{ivano_phase}. They were obtained in a short time period (about 2.5 h) covering an unprecedented broad phase angle range (from $\sim10^{\circ}$ to $\sim155^{\circ}$).  The measured phase functions show a peculiar u-shape with a minimum at a phase angle around $100^{\circ}$. That phase function is not reproduced by any of the $\mu m$-sized randomly oriented particles studied in this work or previously presented in the Amsterdam-Granada Light Scattering Database \citep{munoz_database}. However, laboratory measurements of the phase function of mm-sized particles \citep{munoz_mm-size} show a distinct behavior compared to $\mu m$-sized particle clouds, with a minimum placed at much larger phase angles more in line with the 67P phase curve. Further, recent analysis of the OSIRIS phase functions at low phase angles provides a range of BSE values broader [1.7-3.6] \citep{ivano_BSE} than that based on ground-based observations  [1.7-2.7] \citep{ishiguro}. As shown in Table \ref{tab:table_phase}, the BSE values obtained for our $\mu m$-sized cometary dust analogues are within the range obtained from  ground-based observations. The highest BSE value obtained from the OSIRIS data (3.6) seem to be produced by a cloud of decimetric chunks orbiting 67P nucleus at distances smaller than 100 km \citep{ivano_BSE}. All in all our experimental data suggest that 67P phase function is not dominated by randomly-oriented $\mu m$-sized particles in line with the conclusions of multi-instrument analysis on board Rosetta \citep{carsten}.


  Polarization is a powerful tool to investigate the nature of particles populating cometary comae and asteroidal regoliths. The value of the minimum of the degree of linear polarization together with the slope of polarization curve is generally used to determine the albedo of asteroids.  Moreover, polarimetric criteria are used to classify and refine the asteroid taxonomy \citep{belskaya}. A caveat to direct comparisons between observations and laboratory results is that the assumption of single scattering is no longer valid for such bodies, since they are covered by a layer of regolith. Single scattering is considered to be more adequate to explain the behaviour of low-albedo objects, for which multiple scattering is less important. As a result of multiple scattering the minimum polarization value approaches zero, and the inversion angle $\alpha_0$ decreases. Further, the positive polarization maximum decreases by multiple scattering effects and therefore its slope \citep{shkuratov2004, shkuratov2006, shkuratov2007}. We refer to \cite{kolokolova2015} (chapters 5 and 8) for recent reviews on photometric and polarimetric laboratory measurements of particulate surfaces and individual particles. Taking into account the mentioned multiple scattering effects, we can still retrieve information on the physical properties that determines the observed polarimetric features of asteroidal regolith particles by direct comparison with our experimental data. For instance, in the case of (the still few) Near-Earth Asteroids (NEA) observed so far, the maximum of positive polarization can reach, for low-albedo objects like (3200) Phaethon and (101955) Bennu, much higher values of  linear polarization (up to 40\%), well above the limits reached by the samples analyzed in this work. The covered interval of angles did not include $\alpha_{max}$ but the observed data indicate that $P_{max}$ could occur around $130^{\circ}$. Taking into account the decrease of $P_{max}$ due to multiple scattering effects on the regolith layer, we can expect an even larger $P_{max}$ for the single particles forming the regolith. According to previous experimental \citep{jesus_lunar} and computational \citep{liu_gaussian_sphere} results, a high value of $P_{max}$ shifted toward large phase angles, suggests a surface covered by particles significantly larger than those studied in this work. 

\begin{figure} 
    \includegraphics[width =0.5 
    \textwidth]{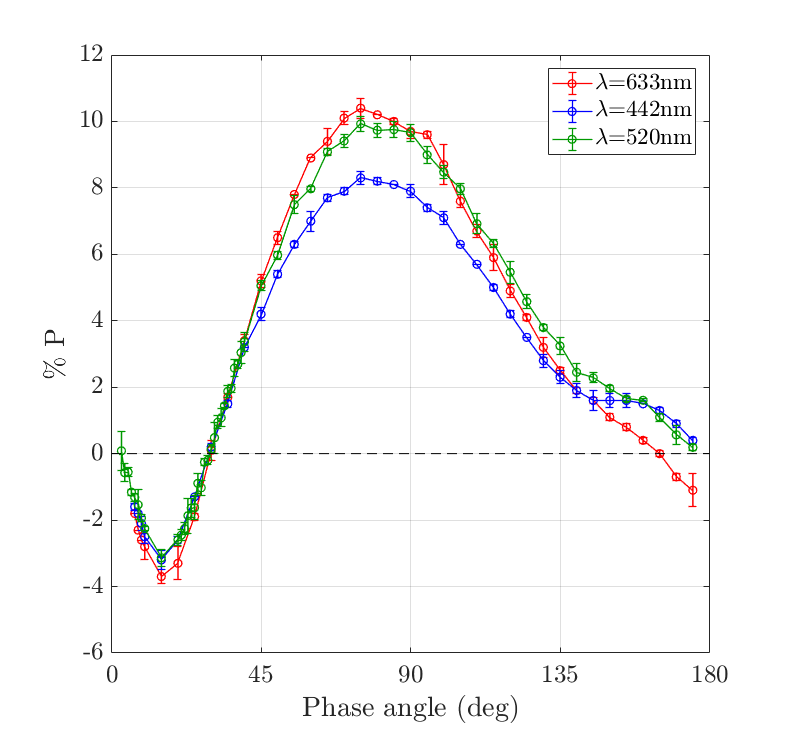}
        \caption{Measured degree of linear polarization curves for the forsterite sample at three different 
        wavelengths.}
    \label{fig:oli_col}
    \end{figure}

\begin{figure} 
     \includegraphics[width =0.5 
     \textwidth]{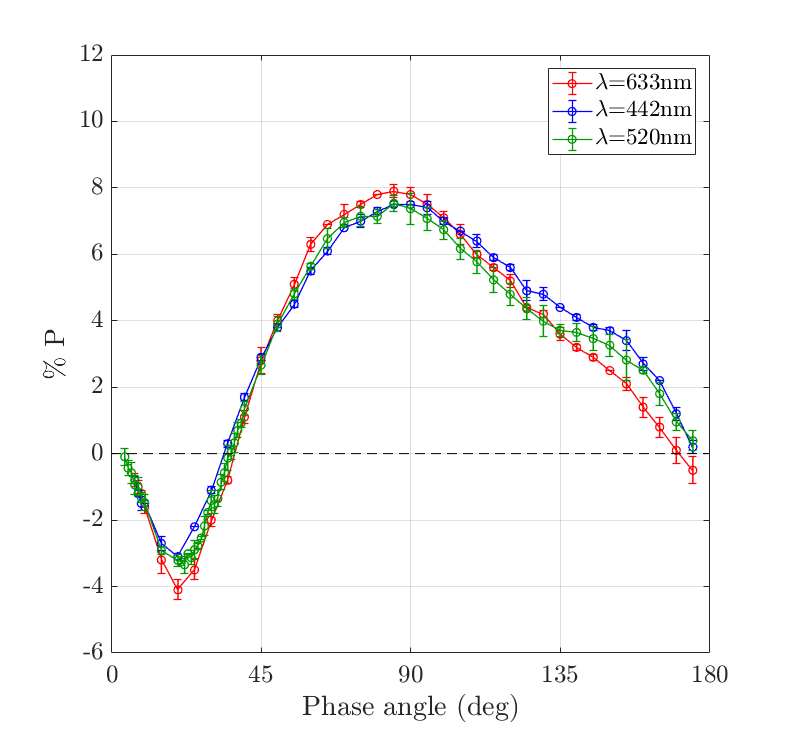}
    \caption{Measured degree of linear polarization curves for the Allende sample at three different wavelengths.}
    \label{fig:all_col}
\end{figure}

    Further, spectropolarimetric data of asteroids show that the gradient of linear polarization for increasing wavelength depends on the taxonomic class and upon the phase angle. In the positive polarization branch, the polarization tends to decrease for increasing wavelength for S-class objects (moderate albedo). Similar dependence of $P_{max}$ with $\lambda$ has been found for single particles that present a significantly higher imaginary part of the refractive index, $k$, at short (488 nm) than at large (647 nm) wavelengths \citep{dabrowska2015}.  Low-albedo objects do just the opposite as is the case of dust samples with a flat (Figures \ref{fig:oli_col}) or moderate dependence of $k$ with the wavelength (\ref{fig:all_col}). Note also that, for any given object of any class, the polarimetric gradient changes sign if the object is observed in the negative or in the positive polarization branch \citep{bagnulo} as is the case of the forsterite and Allende samples studied in this work (Figures \ref{fig:oli_col} and \ref{fig:all_col}).\\

 On the other hand, the angle of inversion of polarization of asteroids is usually around 20$^{\circ}$. A few objects exhibit an inversion angle at lower phase angles ($15^{\circ}-17^{\circ}$). Only a few rare objects, the so-called Barbarians, show high values of inversion angles, up to $28^\circ$ \citep{cellino2006, devogele}. As shown in Section \ref{measurements}, for the size range of the samples presented in this work, the position of the inversion angle depends on sample composition. This is also found by \cite{devogele} in their analysis of Barbarian asteroids. Other observed polarimetric evidence ($P_{min}$ versus inversion angle plot) suggests that these objects should have a very fine surface regolith. This is also in agreement with some estimate of a low thermal inertia for (21) Lutetia, which has also a large inversion angle, not much smaller than that of Barbarians \citep{cellino2016}. Only S-class NEAs objects have $P_{max}$ values (7\% and 10\%) compatible with our measurements. As opposite, there are no asteroids reaching negative polarization deeper than  -2,-3 \%.

The condition of single scattering holds for cometary comae. Observational data show the existence of two classes of comets, based on the degree of linear polarization \citep{levasseur-regourd_class,shestopalov2017}. The high polarization comets, which are considered dust-rich,  and the low polarization comets, which have a higher gas component (Table \ref{tab:table_comet}). The measurements of the samples presented in this work seems to be of the order of the low polarization group. Generally, cometary comae behave qualitatively in a similar manner to our analog particle clouds: they show a negative polarization branch at small phase angles, and a bell-shaped curve at side and backscattering angles, with a maximum of polarization at a phase angle  near 90$^{\circ}$. However, the exact parameters defining the cometary polarization phase curve (see Table \ref{tab:table_comet} for comet 67P in particular), are different from  our laboratory measurements. The 67P polarization phase curve might be shaped by particles larger than the wavelength, but with wavelength-scale surface features that could produce a polarization roughly similar that of a cloud of isolated scatterers. Moreover, the $h$ slope observed for 67P seems to indicate that  it belongs to the family of "high polarization" comets \citep{rosenbush}. This might be in agreement with the depletion of small particles (<1 $\mu$m) observed with Rosetta.
    
Our laboratory data support the interpretation of the decreasing polarization outward the nucleus of 67P, according to which  this results from larger particles near the nucleus and smaller particles moving away from it \citep{rosenbush}.
In order to enable a more consistent comparison between astronomical observations and laboratory data, future studies need to consider more appropriate mixtures of components such as silicates, organics and ices, with improved control on other characteristics such as size, shape  and surface roughness.

\section{Conclusions}

 Our results are summarized as follows:

\begin{itemize}
\item All measured phase functions present the typical behavior of $\mu m$-sized irregular compact dust particles in random orientation. They show strong peaks at large phase angles, almost no structure at side phase angles and soft increase at small phase angles.
\item The measured values of backscattering enhancement (BSE) of our samples are within the range obtained from ground-based observations of cometary comae. According to our experimental data small BSE values seem to indicate a coma population of $\mu m$-sized grains.
\item The dependence of $P_{max}$  with size is opposite for particle sizes belonging to  Rayleigh-resonance ($x_{eff}\approx \lambda$) and  geometric optic ($x_{eff}\gg \lambda$)  regimes. When particles are smaller or of the order of the wavelength (Rayleigh-resonance regimes), $P_{max}$ decreases with size whereas in the geometric optics regime $P_{max}$ tend to increase with size.  
\item For the range of sizes of our samples, the minimum of the negative polarization branch seems to be dependent on both size and composition of the grains whereas the inversion angle value, $\alpha_{0}$, depends on the refractive index.   
\item The measured degree of linear polarization curves are qualitatively similar to those obtained from ground-based observations of cometary comae. This is in line with the finding that the main scatterers in the coma are larger than the wavelength of the incident light, where the surface features of the particles are of the order of the incident wavelength.
 \end{itemize}

\section*{Acknowledgement}
The authors are grateful with the Museo dell'Antartide Felice  Ippolito, Siena for providing the meteorite samples required,  with the Department of Geoscience of the University of Padua for providing the laboratory facilities necessary to produce the samples. 
We are indebted to Roc\'{\i}o M\'arquez from the Scientific Instrumentation center of the University of Granada for providing the SEM images and with Maurizio Gemelli from the Department of Earth Science of the University of Pisa.
This work has been supported by the Plan Nacional de Astronom\'{\i}a y Astrof\'{\i}sica contract AYA2015-67152-R.

\makeatletter
\renewcommand\@biblabel[1]{}
\makeatother

\bibliographystyle{agsm}  
\bibliography{Scattering_of_cometary_analogs_ARXIV}

\end{document}